\documentclass[12pt]{article}

\usepackage[dvips]{graphicx}
\usepackage{amssymb}
\usepackage{amsmath}
\usepackage{array} % to improve presentation of tables
\usepackage{epsfig}
\usepackage{graphics}
\usepackage{colortbl}
\usepackage{hhline}
\usepackage{multirow}

\newcommand{\tsc}[1]{\textsc{#1}}
\newcommand{\tbf}[1]{\textbf{#1}}
\newcommand{\ttt}[1]{\texttt{#1}}

\newcommand{\ExD}{\tsc{ExDiff}}

\newenvironment{Itemize}{\begin{list}{$\bullet$}%
{\setlength{\topsep}{0.2mm}\setlength{\partopsep}{0.2mm}%
\setlength{\itemsep}{0.2mm}\setlength{\parsep}{0.2mm}}}%
{\end{list}}
\newcounter{enumct}

\newenvironment{entry}%
{\begin{list}{}{\setlength{\topsep}{0mm} \setlength{\itemsep}{0mm}
\setlength{\parskip}{0mm} \setlength{\parsep}{0mm}
\setlength{\leftmargin}{20mm} \setlength{\rightmargin}{0mm}
\setlength{\labelwidth}{18mm} \setlength{\labelsep}{2mm}}}%
{\end{list}}
\newenvironment{subentry}%
{\begin{list}{}{\setlength{\topsep}{0mm} \setlength{\itemsep}{0mm}
\setlength{\parskip}{0mm} \setlength{\parsep}{0mm}
\setlength{\leftmargin}{10mm} \setlength{\rightmargin}{0mm}
\setlength{\labelwidth}{18mm} \setlength{\labelsep}{2mm}}}%
{\end{list}}
\newcommand{\itemc}[1]{\item[\textbf{#1}\hfill]}
\newcommand{\iteme}[1]{\item[\texttt{#1}\hfill]}

\newcommand{\drawbox}[1]{\vspace{\baselineskip}\noindent%
\fbox{\texttt{#1}}\vspace{0.5\baselineskip}}

\newcommand{\boxsep}{\vspace{0.5\baselineskip}} 
 
\newlength{\captivewidth}
\newlength{\tablinsep}
\newlength{\halfpagewid}
\newlength{\abstwidth}
\begin{document}
\sloppy
 
\pagestyle{empty}
 
\begin{flushright}
CERN\\
April, 2017
\end{flushright} 
\vskip 1.75cm

\begin{center}
\begin{verbatim}
                ########            #####  
                #                   #    #
                #                   #     #        ###    ###
                #                   #      #      #   #  #   #
                ########  #     #   #      #  #   #      #
                #          #   #    #      #     ###    ###
                #           ###     #     #   #   #      #
                #          #   #    #    #    #   #      #
                ########  #     #   #####     #   #      #
\end{verbatim}
\vspace*{1cm}
%\tbf{{\Huge\bf E}{\Large\bf x}{\Huge\bf D}{\Large\bf iff:}}\\[3mm]
\tbf{{\Large\bf M}{\large\bf onte Carlo generator for Exclusive Diffraction}}\\[3mm]
\tbf{{\large\bf Version 1.0}}\\[10mm]
\tbf{{\large\bf Physics and Manual}} \\[20mm]
{ R.A.~Ryutin, V.A.~Petrov  } \\[7mm]
{\small Institute for High Energy Physics, NRC ``Kurchatov Institute'',} \\
{\small {\it 142 281} Protvino, Russia}

\vskip 1.75cm
{\bf
\mbox{Abstract}}
  \vskip 0.3cm
%---------------Abstract------------------------------------------

\newlength{\qqq}
\settowidth{\qqq}{In the framework of the operator product  expansion, the quark mass dependence of}
\hfill
\noindent
\begin{minipage}{\qqq}

\ExD\ is a Monte Carlo event generator for simulation of 
Exclusive Diffractive processes 
in proton-proton collisions. The present version includes reactions: 
elastic scattering $pp\to pp$ at 7, 8, 13, 14~TeV; 
$pp\to p+R+p$, $R = f_0(1710),\, f_2(1950)$ at 13~TeV. In the future
versions many processes of Central Exclusive Diffractive Production will be added.  

\end{minipage}
\end{center}

%-------------------------------  KEYWORDS  -------------------------------------

\begin{center}
\vskip 0.5cm
{\bf
\mbox{Keywords}}
\vskip 0.3cm

\settowidth{\qqq}{In the framework of the operator product  expansion, the quark mass dependence of}
\hfill
\noindent
\begin{minipage}{\qqq}
Exclusive Diffraction -- Elastic Scattering -- Central Exclusive Diffractive Production -- 
Pomeron-Pomeron -- photon-Pomeron -- proton-proton -- cross sections -- event generator -- forward physics -- Regge-Eikonal model
\end{minipage}

\end{center}

\clearpage

\tableofcontents
 
\clearpage

\pagestyle{plain}
\setcounter{page}{1}
%----------------Introduction---------------------------------------------

\section{Introduction}

% theor. intro

  Substantial  fraction  (about 40\% at LHC)  of  total  cross  section  of  pp  interactions  is  due  to 
diffractive processes. And about 60\% of these events are exclusive. So, simulation of such events
is one of the basic tasks at LHC.

 At this moment there is no unique definition of diffraction: 
\begin{itemize}
\item
Interactions where the beam particles emerge intact or dissociated into low-mass states. 
\item
Interactions mediated by t-channel exchange of object 
with the quantum numbers of the vacuum, color singlet exchange or Pomeron.
Descriptions of the Pomeron are based on phenomenological approaches or on QCD. 
\end{itemize}
In  general  such  processes  lead  to  final  state  particles  separated  by  large 
rapidity gaps. However only a fraction of Large Rapidity Gap (LRG) events is due to  
diffractive processes, gaps can arise also from fluctuations in the 
hadronisation processes.

{\bf The key experimental trigger for diffraction is the angle distribution}, which 
gives the typical diffractive pattern with zero-angle maximum and one 
or, sometimes, two dips. Here wave properties of hadrons play the main role. From 
this distribution we can make the conclusion about size and shape of the 
scatterer, or the "interaction region".  

%% elastic processes and information from them, soft rescattering normalization
{\bf Elastic scattering} is the basic exclusive process, i.e. so called ``standard candle''. Advantages 
of this process are:
\begin{Itemize}
	\item Clear signature: both intitial particles remain intact and should be detected in the final state.
	\item Small number of variables in the differential cross-section.
	\item Large value of the cross-section.
	\item Huge number of experimental data at different energies.
	\item From the theoretical point of view: we have the possibility to extract size and shape of 
	the ``interaction region'' from the slope and fine structure of the $t$-distribution.
	\item Also we can ``calibrate'' diffractive models for further calculations of absorptive corrections in other
	exclusive processes.	
\end{Itemize}
Experimental difficulty is mainly due to closeness of final protons to beams. That is why we need special runs of an accelerator to avoid different contaminations like pile-up events, for example.  

%% Exclusive Central Production, advantages, information from them
%% models, that we use

Another process is the {\bf Central Exclusive Diffractive Production} (CEDP).  CEDP gives us unique experimental 
possibilities for particle searches
and investigations of diffraction itself. This is due to several advantages of the process: 
\begin{Itemize}
	\item clear signature of the process: both protons remain intact plus LRG; 
	\item possibility to use "missing mass method" that improve the mass resolution; 
	\item usually background is strongly suppressed, especially for ``hard'' production, due to $J_z=0$ selection rule; 
	\item spin-parity analysis of the central system can be done by the use of azimuthal distributions; 
	\item interesting measurements concerning the interplay between "soft" and "hard" scales are possible~\cite{diffpatterns},\cite{ryutinEDDE2}.
\end{Itemize}
 All these properties are realized in common CMS/TOTEM detector measurements at LHC~\cite{CTPPSTDR}.

Theoretically calculations of CEDP and elastic scattering are closely related, especially
when we try to take into account all the unitarity effects (absorption, ``gap survival probability'').
% instrumental intro

\begin{figure}[hbt!]
	\begin{center} 
		%\resizebox{0.49\textwidth}{5cm}{%
		\includegraphics[width=0.95\textwidth]{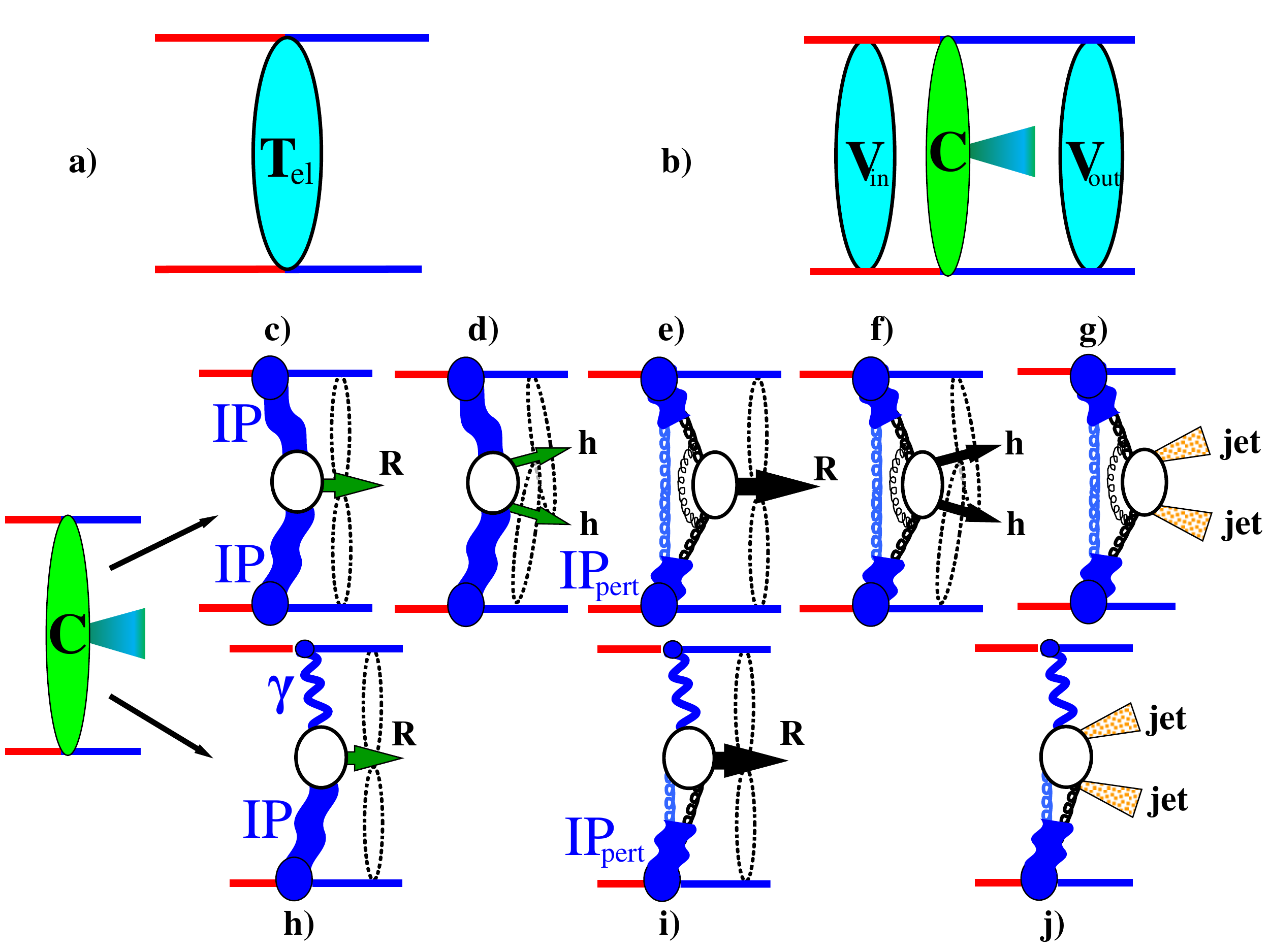}
		%}
		\caption{\label{fig1:difpr} Amplitudes for exclusive diffractive processes: a) elastic scattering with $T_{el}$ presented in~(\ref{eq:Telt}); b) general process of central exclusive diffractive production (CEDP) with central production born amplitude $C$ and absorptive corrections $V_{in}=V(s,b)$ and $V_{out}=V(s^{\prime},b)$ presented in~(\ref{MUgen}),(\ref{Vblobs}). Versions of the CEDP Born amplitudes are: c) low mass resonance production; d) di-hadron production; e) high mass resonance production; f) high mass di-hadron production; g) di-jet production; h) nonperturbative photon-pomeron low mass resonance production; i) perturbative photon-pomeron high mass resonance production and j) high mass di-jet production. Possible additional corrections due to unitarization procedure are depicted as dotted ovals. In d) and f) it is possible to produce also di-bosons like $\gamma\gamma$, $ZZ$ or $WW$ instead of di-hadrons $hh$.}
	\end{center} 
\end{figure}

In the paper we present a new  Monte-Carlo event generator \ExD. The 
generator is devoted to the simulation of 
Exclusive Diffractive processes in proton-proton collisions (elastic process and CEP of low mass resonances 
in this version). All these processes are depicted in the Fig.~\ref{fig1:difpr}.

The present version includes reactions: 
$pp\to pp$ at 7, 8, 13, 14~TeV; $pp\to p+R+p$, $R = f_0(1710),\, f_2(1950)$ 
at 13~TeV. In the future versions many other processes of CEP will be 
added: $pp\to p+R+p$, where $R$ is a resonance like Higgs boson, 
for example; $pp\to p+A+B+p$, where $A$, $B$ are hadrons, jets or gauge bosons (see Figs.~\ref{fig1:difpr}(c)-(j)). 

%%%%%%%%%%%%%%%%%%%%%%%%%%%%%%%%%%%

\newpage
%---------------main part---------------------------------------

\section{Physics Overview}
\label{s:physics}

\subsection{Elastic scattering}
\label{ss:elastic}

\subsubsection{Kinematics}
\label{sss:elastickin}

The diagram of the elastic scattering $p+p\to p+p$ is presented in 
Fig.~\ref{fig1:difpr}. 
The momenta are $p_1$, $p_2$, $p_1^{\prime}$, $p_2^{\prime}$ respectively. 
In the center-of-mass frame these can be represented as follows 
($p\equiv(p_0,p_z,\vec{p}\,)$, $\vec{p}\equiv(p_x,p_y)$):
 \begin{equation}
 \label{kin1a}p_1=\left(\frac{\sqrt{s}}{2},\frac{\sqrt{s}}{2}\beta ,
 \vec{0}\right),\; 
 p_2=\left(\frac{\sqrt{s}}{2},-\frac{\sqrt{s}}{2}\beta ,\vec{0}\right).
\end{equation}
With this notation, the four momentum transfer is
\begin{eqnarray}
&&\label{kin1c}\Delta=\left( 0,
\frac{-t}{\sqrt{s}\beta} ,
\vec{\Delta}
\right),\,
p_1^{\prime}=p_1-\Delta,\,
p_2^{\prime}=p_2+\Delta,\\
&&\label{kin1e}
-t=\tau^2=\frac{s\beta^2}{2}\left( 
1-\sqrt{1-\frac{4\vec{\Delta}^{2}}{s\beta^2}}
\right)
\simeq\vec{\Delta}^{2},\,
\beta=\sqrt{1-\frac{4m_p^2}{s}}.
 \end{eqnarray}

Cross-section for elastic scattering can be calculated as
\begin{eqnarray}
\label{diffsech}
&& \label{eq:dsigeldt}\frac{d\sigma_{el}}{d\tau^2} = \frac{|T_{el}(s,\tau)|^2}{16\pi s^2}\,,\\
&& \label{eq:Telt}T_{el}(s,\tau) = 4\pi s\int_0^{\infty}db^2J_0(b\tau) T_{el}(s,b).
\end{eqnarray}

\subsubsection{Model for the elastic scattering}
\label{sss:elasticmodel}
 
 Here is the outlook of the model for elastic scattering used in this version of the generator. 
 
In the Regge-eikonal approach the elastic nonflip scattering amplitude looks like 
\begin{eqnarray}
&&\label{eq:TelbEik} T_{el}(s,b) = \frac{e^{2\mathrm{i}\delta_{el}(s,b)}-1}{2i}\,,\\
&&\label{eq:deltaelbEik} \delta_{el}(s,b) = \frac{1}{16\pi s}\int_0^{\infty}d\tau^2 J_0(b\tau)\delta_{el}(s,\tau)\,,
\end{eqnarray}
where $s$ and $t=-\tau^2$ are the Mandelstam variables, $b=|\vec{b}|$ is the impact parameter, and eikonal $\delta(s,\tau)$ is
parametrized as in the work~\cite{godizovElastic}
\begin{equation}
\label{eq:deltaeltEikG} \delta_{el}(s,\tau)=\;g^2_{pp\rm P}(-\tau^2)
\left(i+{\rm tan}\frac{\pi(\alpha_{\rm P}(-\tau^2)-1)}{2}\right)\pi\alpha'_{\rm P}(-\tau^2)\left(\frac{s}{2s_0}\right)^{\alpha_{\rm P}(-\tau^2)}\,,
\end{equation}
where
\begin{equation}
\alpha_{\rm P}(t) = 1+\frac{\alpha_{\rm P}(0)-1}{1-\frac{t}{\tau_a}}\,,\;\;\;\;g_{pp\rm P}(t)=\frac{g_{pp\rm P}(0)}{(1-a_gt)^2}\,,
\end{equation} 
\begin{table}[ht]
	\begin{center}
		\begin{tabular}{|l|l|}
			\hline
			\bf Parameter          & \bf Value                   \\
			\hline
			$\alpha_{\rm P}(0)-1$  & $0.109$            \\
			$\tau_a$               & $0.535$ GeV$^2$  \\
			$g_{pp\rm P}(0)$       & $13.8$ GeV         \\
			$a_g$                  & $0.23$ GeV$^{-2}$ \\
			\hline
		\end{tabular}
	\end{center}
\end{table}

\subsection{Central Exclusive Diffractive Production}
\label{ss:CEDP}

\subsubsection{Kinematics of CEDP}
\label{sss:CEDPkin}

Let us consider the kinematics of two CEDP processes
\begin{eqnarray}
&&\label{eddeRprod} h_1(p_1)+h_2(p_2)\to h_1(p_1^{\prime})+R(p_R)+h_2(p_2^{\prime}),\\
&&\label{eddeABprod} h_1(p_1)+h_2(p_2)\to h_1(p_1^{\prime})+\{a(k_a)+b(k_b)\}+h_2(p_2^{\prime}),
\end{eqnarray}
with four-momenta indicated in parentheses. Initial hadrons remain 
intact, $\{a\; b\}$ can be di-boson or di-hadron system and R denotes 
a resonance, ``+'' signs denote LRG. 

We use the following set of variables:
\begin{eqnarray}
s&=&(p_1+p_2)^2,\; s^{\prime}=(p_1^{\prime}+p_2^{\prime})^2,\; t_{1,2}=(p_{1,2}-p_{1,2}^{\prime})^2,\nonumber\\
s_{1,2}&=&(p_{1,2}^{\prime}+p_R)^2\; \mbox{\rm or}\; (p_{1,2}^{\prime}+k_a+k_b)^2,
\end{eqnarray}
\begin{eqnarray}
s_{1\{a,b\}}&=&(p_{1}^{\prime}+k_{a,b})^2,\; s_{2\{a,b\}}=(p_{2}^{\prime}+k_{a,b})^2,\nonumber\\
\hat{t}_{a,b}&=&(p_1-p_1^{\prime}-k_{a,b})^2=(p_2-p_2^{\prime}-k_{b,a})^2,\nonumber\\
\bar{s}&=&\frac{s-2m^2}{2}+\frac{s}{2}\sqrt{1-\frac{4m^2}{s}}\simeq s.\label{kin:invars}
\end{eqnarray}
In the light-cone representation $p=\{p_+,p_-; \vec{p}_{\perp}\}$
\begin{eqnarray}
&&\hspace*{-0.3cm}p_1\!=\!\left\{ \sqrt{\frac{\bar{s}}{2}},\frac{m^2}{\sqrt{2\bar{s}}};\; \vec{0}\right\},\;\!\!
\Delta_1\!=\!\left\{ 
\xi_1\sqrt{\frac{\bar{s}}{2}},\frac{-\vec{\Delta}_1^2-\xi_1 m^2}{(1-\xi_1)\sqrt{2\bar{s}}};\; \vec{\Delta}_1
\right\},\;\nonumber\\
&&\hspace*{-0.3cm}p_2\!=\!\left\{ \frac{m^2}{\sqrt{2\bar{s}}},\sqrt{\frac{\bar{s}}{2}};\; \vec{0}\right\},\;\!\!
\Delta_2\!=\!\left\{
\frac{-\vec{\Delta}_2^2-\xi_2 m^2}{(1-\xi_2)\sqrt{2\bar{s}}},\xi_2\sqrt{\frac{\bar{s}}{2}};\; \vec{\Delta}_2
\right\}\nonumber\\
&&\hspace*{-0.3cm}p_{1,2}^{\prime}=p_{1,2}-\Delta_{1,2},\;
p_{1,2}^2=p_{1,2}^{\prime\; 2}=m^2, \label{kin:momenta22}
\end{eqnarray}
and additional notations for the $2\to 4$ process are
\begin{eqnarray}
k_{a,b}&\simeq&\left\{\frac{m_{\perp}}{\sqrt{2}}\mathrm{e}^{y\mp\eta},\frac{m_{\perp}}{\sqrt{2}}\mathrm{e}^{-y\pm\eta};\; \pm\vec{k} 
\right\},\nonumber\\
k_{a,b}^2&=&m_0^2,\; m_{\perp}^2=m_0^2+\vec{k}^2. \label{kin:momenta23}
\end{eqnarray}
Here $\xi_{1,2}$ are fractions of hadrons' longitudinal momenta lost, $y$ denotes the rapidity of the central system, $\eta=(\eta_b-\eta_a)/2$, where $\eta_{a,b}$ are rapidities of particles $a,b$. From the above notations we can
obtain the relations:
\begin{eqnarray}
t_{1,2}&=&\Delta_{1,2}^2 \simeq -\frac{\vec{\Delta}_{1,2}^2+\xi_{1,2}^2m^2}{1-\xi_{1,2}}\;\;\simeq
-\vec{\Delta}_{1,2}^2,\;\xi_{1,2}\to 0\nonumber\\
s_{1,2}&\simeq&\xi_{2,1}s,\, \tau_{1,2}=\sqrt{-t_{1,2}}, \nonumber\\
M^2&=& (\Delta_1+\Delta_2)^2\simeq \xi_1\xi_2s+t_1+t_2-2\sqrt{t_1t_2}\cos\phi_0\nonumber\\
M_{\perp}^2&=& \xi_1\xi_2s \simeq M^2+|t_1|+|t_2|+2\sqrt{t_1t_2}\cos\phi_0\nonumber\\
\cos\phi_0&=&\frac{\vec{\Delta}_1\vec{\Delta}_2}{|\vec{\Delta}_1||\vec{\Delta}_2|}\label{kin:relations}
\end{eqnarray}
and additional set for the $2\to 4$ process
\begin{eqnarray}
s_{1\{a,b\}}&\simeq& m^2(1-\xi_1)^2+m_0^2+
\frac{M}{M_{\perp}}\frac{s_1}{2}(1\pm\tanh\eta)(1-\xi_1),\nonumber\\
s_{2\{a,b\}}&\simeq& m^2(1-\xi_2)^2+m_0^2+
\frac{M}{M_{\perp}}\frac{s_2}{2}(1\mp\tanh\eta)(1-\xi_2),\nonumber\\
\hat{t}_{a,b}&\simeq& m_0^2-\frac{M M_{\perp}}{2}(1\pm\tanh\eta),\nonumber\\
\label{kin:relations2}m_{\perp}&\simeq&\frac{M_{\perp}}{2\cosh\eta}.
\end{eqnarray}

Physical region of diffractive events with two large rapidity gaps 
is defined by the following  
kinematical cuts:
\begin{eqnarray}
\label{eq:tlimits}
&&0.01\; GeV^2\le |t_{1,2}|\le\; \sim 1\; GeV^2\;{,} \\
&&\label{eq:xilimits}
\xi_{min}\simeq\frac{M^2}{s \xi_{max}}\le \xi_{1,2}\le \xi_{max}\sim 0.1\;,\\
\label{eq:kappalimits}
&&\left(\sqrt{-t_1}-\sqrt{-t_2}\right)^2\le\kappa\le\left(\sqrt{-t_1}+\sqrt{-t_2}\right)^2\\
&&\kappa=\xi_1\xi_2s-M^2\ll M^2\nonumber
\end{eqnarray}
We can write the relations in terms of $y_{1,2}$ (rapidities 
of hadrons) and $y$. For instance:
\begin{eqnarray}
&&\xi_{1,2}\simeq\frac{M_{\perp}}{\sqrt{s}}\mathrm{e}^{\pm y},\;
|y|\le y_0=\ln\left(\frac{\sqrt{s}\xi_{max}}{M}\right),\nonumber\\
\label{eq:raplimits}&&|y_{1,2}|=\frac{1}{2}\ln\frac{(1-\xi_{1,2})^2s}{m^2-t_{1,2}},\nonumber\\
&& |y|\le 6.5,\; |y_{1,2}|\ge 8.75 \mbox{ for } \sqrt{s}=7\;\mathrm{TeV},\nonumber\\
&&|\tanh\eta|\le\sqrt{1-\frac{4m_0^2}{M^2}}.
\end{eqnarray}
%$y_0\simeq 2.5$ for $\sqrt{s}=14\; TeV$ and $M=O(100\; GeV)$.

Differential cross-sections for the above processes can be represented as
\begin{eqnarray}
&& \frac{d\sigma^{\mbox{\tiny CEDP}}_{R}}{d\vec{\Delta}_1^2d\vec{\Delta}_2^2d\phi_0 dy}\simeq\frac{\left| T^{\mbox{\tiny CEDP}}_R\right|^2}{2^8\pi^4ss^{\prime}},\label{eq:EDDEcsGENR}\\
&& \frac{d\sigma^{\mbox{\tiny CEDP}}_{ab}}{d\vec{\Delta}_1^2 d\vec{\Delta}_2^2 d\phi_0 dy dM^2 d\Phi_{ab}}\simeq
\frac{\left| T^{\mbox{\tiny CEDP}}_{ab}\right|^2}{2^{9}\pi^5 ss^{\prime}},\nonumber\\
&& d\Phi_{ab}=\frac{d^4k_a}{(2\pi)^2} \delta(k_a^2-m_0^2) \delta(k_b^2-m_0^2)\nonumber\\
&&\;\;= \frac{d\vec{k}^2}{8\pi M^2 \sqrt{1-\frac{4(\vec{k}^2+m_0^2)}{M^2}}}=\frac{d\eta}{16\pi\cosh^2\eta}\label{eq:EDDEcsGENab}.
\end{eqnarray}

$T^{\mbox{\tiny CEDP}}$ is given 
by the following analytical
expression:
\begin{eqnarray}
&&{T}^{\mbox{\tiny CEDP}}(p_1, p_2, \Delta_1, \Delta_2) = \int \frac{d^2\vec{q}_T}{(2\pi)^2} \,
\frac{d^2\vec{q}^{\;\prime}_T}{(2\pi)^2} \; V(s, \vec{q}_T) \;
\nonumber \\
&&\label{MUgen} \times {C}( p_1-q_T, p_2+q_T,\Delta_{1T}, \Delta_{2T}) \,
V(s^{\prime}, \vec{q}^{\;\prime}_T) \;,
\\
&&\label{Vblobs} V(s, \vec{q}_T) = \int d^2\vec{b} \, {\mathrm e}^{i\vec{q}_T
	\vec{b}} V(s,b),\,\,\,\,\,
V(s,b)=\sqrt{1+2\mathrm{i} T_{el}(s, b)}
\end{eqnarray}
where $\Delta_{1T} = \Delta_{1} -q_T - q^{\prime}_T$, $\Delta_{2T}
= \Delta_{2} + q_T + q^{\prime}_T$, ${C}$ is the ``bare'' amplitude of the
process $p+p\to p+X+p$. 

\subsubsection{Model for CEDP}
\label{sss:CEDPmodel}

In the case of the eikonal representation of the
elastic amplitude $T_{el}$ we have
\begin{equation}
\label{eq:Veik}
V(s, \vec{q}_T) = \int d^2\vec{b} \, \mathrm{e}^{\mathrm{i}\vec{q}_T
	\vec{b}} \mathrm{e}^{\mathrm{i}\delta_{el}(s,b)}.
\end{equation}
In this case
amplitude~(\ref{MUgen}) can be rewritten as
\begin{eqnarray}
T^{\mbox{\tiny CEDP}}(\vec{\Delta}_1, \vec{\Delta}_2) &=&
\int \frac{d^2\vec{b}}{2\pi} \, {\mathrm e}^{
	-{\mathrm i}\vec{\delta}\vec{b}-\Omega(s,b)-\Omega(s^{\prime},b)
}\times\nonumber\\
&& 
\int \frac{d^2\vec{\kappa}}{2\pi} \, {\mathrm e}^{{\mathrm i}\vec{\kappa}\vec{b}}
\,{C}(\vec{\Delta}-\vec{\kappa}, \vec{\Delta}+\vec{\kappa}),\nonumber\\
\Omega(s,b)&=&-\mathrm{i}\delta_{el}(s,b),\nonumber\\
\vec{\Delta}&=&\frac{\vec{\Delta}_2+\vec{\Delta}_1}{2},\; 
\vec{\delta}=\frac{\vec{\Delta}_2-\vec{\Delta}_1}{2},\,
\label{MUeik}\vec{\kappa}=\vec{\delta}+\vec{q}_T+\vec{q}^{\;\prime}_T.
\end{eqnarray}

Calculations for concrete expressions of $C$ in the general case were considered in~\cite{ryutinEDDE2},\cite{ryutinEDDE1}.

Here we present the model for resonances in CEDP based on the schemes for single diffractive dissociation~\cite{godizovCEDPRes} and tensor resonance decays~\cite{godizovSD}. Absorptive corrections are calculated according to~\cite{godizovCEDPRes}.
\begin{eqnarray}
&& \!\!\!\!\!\!\!\!\!\!\! C(t_1,t_2,\xi_1,\xi_2) = \pi^2 g_{\rm P\rm P f}(0,0)
\left\{ \prod_{n=1}^2 g_{pp\rm P}(t_n)\alpha'_{\rm P}(t_n)\left({\mathrm i}+{\rm tan}\frac{\pi(\alpha_{\rm P}(t_n)-1)}{2}\right)
\left(\frac{1}{\xi_n}\right)^{\alpha_{\rm P}(t_n)}\right\}\eta_J\nonumber\\
&& \!\!\!\!\!\!\!\!\!\!\! \eta_0 = 1,\, \eta_1 =  \Delta_1^\mu\,e^{(\lambda)}_\mu(\Delta_1+\Delta_2),\,
\eta_2 = \Delta_1^\mu\Delta_1^\nu\,e^{(\lambda)}_{\mu\nu}(\Delta_1+\Delta_2)
\label{eq:CEDPG}
\end{eqnarray}
where $J$ is the spin of a central particle, and 
$e^{(\lambda)}(\Delta_1+\Delta_2)$ are states with the helicity $\lambda$.

\newpage

\section{Program Overview}
\label{s:programoverview}

\subsection{General information}
\label{ss:outlook} 

\begin{figure}[hbt!]
	\begin{center} 
		%\resizebox{0.49\textwidth}{5cm}{%
		\includegraphics[width=0.8\textwidth]{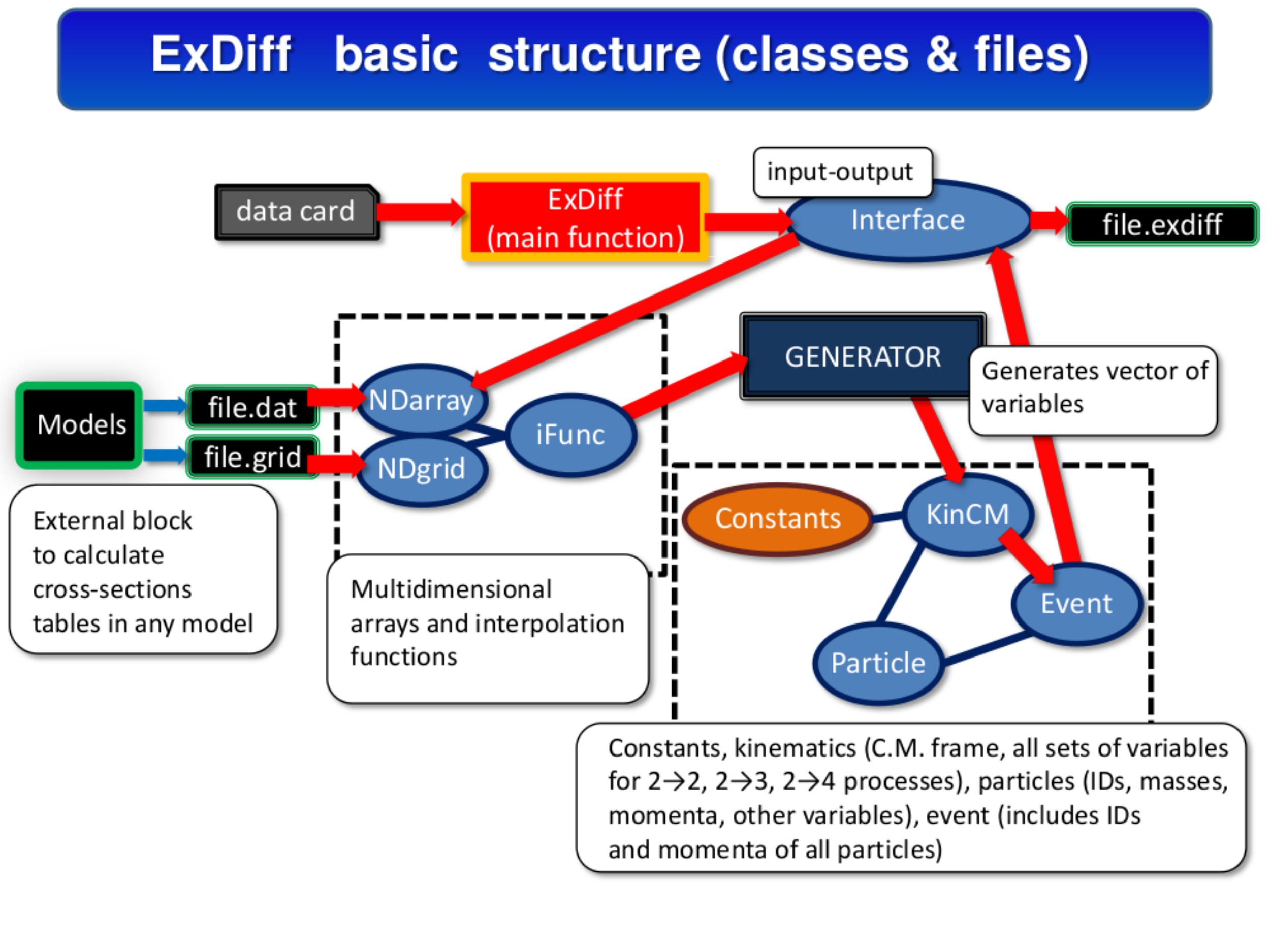}
		%}
		\caption{\label{fig2:structure} General structure of the generator. Classes, subroutines and files.}
	\end{center} 
\end{figure}

\ExD\ is written in a modular way using C++. The general structure of the generator is shown on the 
Fig.~\ref{fig2:structure}. In this subsection there is an outlook. You can 
find more detailed description of all the elements in next subsections.

\subsubsection{Data card}
\label{sss:datacard}
The first element is the data card. Let us consider an example\\ \texttt{../ExDiff1.0/config/card\_sample.card}: 
\begin{verbatim}
1 0 0 1 1000 0 0
===========================================================================
IDauthors IDprocess IDenergy version Nevents IDinput_format IDoutput_format
===========================================================================
\end{verbatim}

Only the first line with \texttt{int} type of numbers is used. Sequence of parameters of the first string with possible values for \ExD\ 1.0:\\

\begin{itemize}
\item \texttt{IDauthors} defines authors of the model used for a process:\\
\texttt{IDauthors=1}: A. Godizov (only one for this version).
\item \texttt{IDprocess} defines the process:\\
\texttt{IDprocess=0}: elastic $p+p\to p+p$ scattering;\\
\texttt{IDprocess=1}: CEDP of low mass resonance $f_0(1710)$, $p+p\to p+f_0(1710)+p$;\\
\texttt{IDprocess=2}: CEDP of low mass resonance $f_2(1950)$, $p+p\to p+f_2(1950)+p$;\\
\item \texttt{IDenergy} defines the collision energy of the process:\\
\texttt{IDenergy=0}: $13$~TeV; (for all processes in this version)\\
\texttt{IDenergy=2}: $7$~TeV; (only for elastic in this version)\\
\texttt{IDenergy=3}: $8$~TeV; (only for elastic in this version)\\
\texttt{IDenergy=4}: $14$~TeV; (only for elastic in this version)\\
\item \texttt{version} is auxiliary parameter that is used to define different versions. Default value is $1$.
\item \texttt{Nevents} defines number of events to generate.
\item \texttt{IDinput\_format} defines a set of variables for kinematics. Default value is $0$. It means
that we use the following sets of variables: $\{t,\phi\}$ (elastic scattering); $\{ \tau_1,\tau_2,\phi_0,\ln\xi_1,\phi_1\}$ (CED resonance production); $\{ \tau_1,\tau_2,\phi_0,\ln\xi_1,\phi_1,\eta,\phi_a\}$ (CED di-jet, di-hadron, di-boson production).
\item \texttt{IDoutput\_format} defines output format. Default value is $0$. In \ExD\ v1.0 the simple
file \texttt{../ExDiff1.0/output/A1M1E0\_1F0.exdiff} generated, where the output looks like
\begin{verbatim}
============================== ExDiff Event ==================================
------------------------------------------------------------------------------
ID            E           px            py              pz
------------------------------------------------------------------------------
2212  6.50000000e+03	 0.00000000e+00	 0.00000000e+00	 6.49999993e+03
2212  6.50000000e+03	 0.00000000e+00	 0.00000000e+00	-6.49999993e+03
------------------------------------------------------------------------------
2212  6.49922495e+03	 1.20107011e-01	 1.07105789e-01	 6.49922488e+03
10331 1.72153048e+00	-3.95672633e-02	-9.27803302e-02	-1.71445119e-01
2212  6.49905351e+03	-8.05397481e-02	-1.43254592e-02	-6.49905344e+03
==============================================================================
============================== ExDiff Event ==================================
------------------------------------------------------------------------------
ID            E           px            py              pz
------------------------------------------------------------------------------
2212  6.50000000e+03	 0.00000000e+00	 0.00000000e+00	 6.49999993e+03
2212  6.50000000e+03	 0.00000000e+00	 0.00000000e+00	-6.49999993e+03
------------------------------------------------------------------------------
2212  6.49922495e+03	 1.20107011e-01	 1.07105789e-01	 6.49922488e+03
10331 1.72153048e+00	-3.95672633e-02	-9.27803302e-02	-1.71445119e-01
2212  6.49905351e+03	-8.05397481e-02	-1.43254592e-02	-6.49905344e+03
==============================================================================
...
\end{verbatim}
where the first column containes particles IDs (standard PDG numbering scheme~\cite{PDGnumpart}), and
other four columns contain their four-momentum. Output filename is constructed from 
parameters in the data card:\\
\\
\texttt{A<IDauthor>M<IDprocess>E<IDenergy>\_<version>F<IDinput\_format>.exdiff}  
\end{itemize}

\subsubsection{Block of models}
\label{sss:models}

This block contains data files (simple txt files) obtained somehow from
external calculations. Naming of these files is the same as for 
\texttt{*.exdiff}. 

The file \texttt{../ExDiff1.0/modeldata/*.grid} contains
one column with values of kinematical variables 
that author use to parametrize differential cross section
of the process under consideration. For each variable \texttt{v\_i} we have the interval
divided by \texttt{Nv\_i} parts. The file looks like
\begin{verbatim}
1rst value of v_1
2nd value of v_1
...
(Nv_1+1)th value of v_1
1rst value of v_2
...
(Nv_2+1)th value of v_2
1rst value of v_3
...
(Nv_3+1)th value of v_3
...
(Nv_n+1)th value of v_n
\end{verbatim}
Values are of the \texttt{double} type. \texttt{n} is the number of variables in the set.

By default sets of variables for different kinematics are
\begin{itemize}
\item $2\to 2$ process: $\{t,\phi\}$; 
\item $2\to 3$ process: $\{ \tau_1,\tau_2,\phi_0,\ln\xi_1,\phi_1\}$;
\item $2\to 4$ process: $\{ \tau_1,\tau_2,\phi_0,\ln\xi_1,\phi_1,\eta,\phi_a\}$.
\end{itemize}
$\phi$, $\phi_1$, $\phi_a$ variables are uniformly distributed in the interval $0\to2\pi$, that is why
their values are simply generated as $2\pi\times$(random number in the $[0,1]$ interval). So, finally 
we have one significant variable for $2\to 2$ process, four variables for $2\to3$ process 
and five variables for $2\to4$ process.
		
The file \texttt{../ExDiff1.0/modeldata/*.dat} contains
one column with values of the cross section of the process. This file is loaded to
the multidimensional array which corresponds to the class \ttt{ExDiff::NDarray} (see below).

\subsection{Basic classes and functions}
\label{ss:classes}

Here we describe only basic classes and some of their methods and variables which can be used by developers.

\drawbox{ExDiff::NDarray}\label{c:NDarray}
\begin{entry}
\itemc{Purpose:} to initialize multidimensional dynamical array of any type or class.
\itemc{ }
\itemc{Initialization example:}
\itemc{} ...
\itemc{} \ttt{std::vector<std::size\_t> dim = \{129,129,9,9\};}
\itemc{} \ttt{ExDiff::NDarray<double> dat(dim)};
\itemc{} ...
\itemc{}
\itemc{} Here we have vector \ttt{dim} which defines number of node points for each variable. Indexes are 
$0\to128$ for the first variable, the same for the second one and so on.
\itemc{}
\itemc{Methods:}
\itemc{ }

\iteme{T\& get(const std::vector<std::size\_t>\& indexes)}\label{m:NDarray.get} 
\itemc{ } gets the value of the cross section (\ttt{T} is \ttt{double} in examples) in the node point 
defined by the vector of indexes of variables. {\bf Example:}
\itemc{ }
\itemc{ } \ttt{std::vector<std::size\_t> indexes = \{1,2,3,4\};}
\itemc{ } \ttt{double cs = dat.get(indexes);}
\itemc{ }

\iteme{std::vector<size\_t>\& GetDimensions()}\label{m:NDarray.GetDimensions} 
\itemc{ } gets the value of the private \ttt{dimensions} vector, which is defined by the vector \ttt{dim} in the 
initialization example above. {\bf Example:}
\itemc{ }
\itemc{ } \ttt{std::vector<std::size\_t> gdim = dat.GetDimensions();}
\itemc{ }

\iteme{std::vector<T>\& GetValues()}\label{m:NDarray.GetValues} 
\itemc{ } gets the value of the private \ttt{values} vector of any type or class \ttt{T}, which is filled by numbers loaded from \ttt{*.dat} file or calculated in some other way. {\bf Example:}
\itemc{ }
\itemc{ } \ttt{std::vector<double> gvalues = dat.GetValues();}
\itemc{ }

\iteme{std::size\_t computeTotalSize(const std::vector<std::size\_t>\& dimensions\_)}\label{m:NDarray.computeTotalSize} 
\itemc{ } calculates the total size of the array by multiplication of all dimensions.
\itemc{ }

\iteme{std::size\_t computeIndex(const std::vector<std::size\_t>\& indexes)}\label{m:NDarray.computeIndex} 
\itemc{ } calculates the global index of the array from the input node point \ttt{indexes} to read the value
of the cross section from the vector \ttt{values}, i.e. converts vector of indexes to the single index. {\bf Example:} (used by \ttt{get})
\itemc{ }
\itemc{ } \ttt{return values[computeIndex(indexes)];}
\itemc{ }

\iteme{std::vector<std::size\_t> computeIndexes(std::size\_t index)}\label{m:NDarray.computeIndexes} 
\itemc{ } converts back the single index to the vector of indexes of the node point.
\itemc{ }

\iteme{printC()}\label{m:NDarray.printC} 
\itemc{ } prints the general information about the class object.
\itemc{ }

\iteme{LoadFromFile(std::string filename)}\label{m:NDarray.LoadFromFile} 
\itemc{ } loads data from a \ttt{*.dat} file after the initialization.
\itemc{ }

\itemc{Variables:}
\itemc{ }

\iteme{std::vector<std::size\_t> dimensions ({\it input})  :}\label{v:NDarray.dimensions}  
private vector, which contains dimensions of the array (numbers of node points for each variable).
\itemc{ }

\iteme{std::vector<T> values ({\it input})  :}\label{v:NDarray.values}  
private vector, which contains values of any class or type \ttt{T} loaded from a 
file or calculated in some other way.
\itemc{ }

\itemc{Used by:} \ttt{iFunc}, \ttt{Interface}.
\itemc{Includes:} 
\ttt{<cmath>}
\ttt{<complex>}
\ttt{<cstdlib>}
\ttt{<iostream>}
\ttt{<fstream>}
\ttt{<map>}
\ttt{<string>}
\ttt{<utility>}
\ttt{<vector>}
\ttt{<cassert>}
\end{entry}

\drawbox{ExDiff::NDgrid}\label{c:NDgrid}
\begin{entry}
	\itemc{Purpose:} to initialize special array which contains node points for each variable in the set (from \ttt{*.grid} file or by the use of some calculations). 
\itemc{ }
\itemc{Initialization example:}
\itemc{} ...
\itemc{} \ttt{std::vector<std::size\_t> dim = \{129,129,9,9\};}
\itemc{} \ttt{ExDiff::NDgrid<double> grid(dim)};
\itemc{} ...
\itemc{}
\itemc{} Here we have vector \ttt{dim} which defines numbers of node points for each variable. Indexes are 
$0\to128$ for the first variable, the same for the second one and so on.
\itemc{}
\itemc{Methods:}

\itemc{ }
\iteme{T\& GetVar(const std::size\_t\& varindex, const std::size\_t\& index)}\label{m:NDgrid.GetVar} 
\itemc{ } gets the value of a variable (type \ttt{T} is double as in examples above) with the 
number \ttt{varindex} (in the set of variables) in the node point with the number \ttt{index}. Use this method
only to change node points of the grid inside an object of the class.
\itemc{ }

\itemc{ }
\iteme{std::vector<size\_t>\& GetDimensions()}\label{m:NDgrid.GetDimensions} 
\itemc{ } gets the value of the private \ttt{dimensions} vector, which is defined by the vector \ttt{dim} in the 
initialization example above. {\bf Example:}
\itemc{ }
\itemc{ } \ttt{std::vector<std::size\_t> gdim = grid.GetDimensions();}
\itemc{ }

\iteme{std::vector<T>\& GetValues()}\label{m:NDgrid.GetValues} 
\itemc{ } gets the value of the private \ttt{values} vector of any type or class \ttt{T}, which is filled by numbers loaded from \ttt{*.grid} file or calculated in some other way. {\bf Example:}
\itemc{ }
\itemc{ } \ttt{std::vector<double> gvalues = grid.GetValues();}
\itemc{ }

\iteme{std::vector<T>\& get(const std::vector<std::size\_t>\& \_indexes)}\label{m:NDgrid.get} 
\itemc{ } gets corresponding vector node point (values of each variable are of the type or class \ttt{T}).
\itemc{ }

\iteme{std::size\_t computeIndex(const std::size\_t\& varindex, }
\iteme{const std::size\_t\& index)}
\label{m:NDgrid.computeIndex} 
\itemc{ } computes the global index of the concrete variable (with the number \ttt{varindex} in the set) 
node point (with the node number \ttt{index} for this variable) in the \ttt{NDgrid} vector.
\itemc{ }

\iteme{std::size\_t computeTotalSize(const std::vector<std::size\_t>\& \_dimensions)}\label{m:NDgrid.computeTotalSize} 
\itemc{ } calculates the total size of the array by summation of all dimensions.
\itemc{ }

\iteme{printC()}\label{m:NDgrid.printC} 
\itemc{ } prints the general information about the class object.
\itemc{ }

\iteme{LoadFromFile(std::string filename)}\label{m:NDgrid.LoadFromFile} 
\itemc{ } loads data from a \ttt{*.grid} file after the initialization.
\itemc{ }

\itemc{Variables:}
\itemc{ }

\iteme{std::vector<std::size\_t> dimensions ({\it input})  :}\label{v:NDgrid.dimensions}  
private vector, which contains dimensions of the array (numbers of node points for each variable).
\itemc{ }

\iteme{std::vector<T> values ({\it input})  :}\label{v:NDgrid.values}  
private vector, which contains values of any class or type \ttt{T} loaded from a 
file or calculated in some other way.
\itemc{ }

\itemc{Used by:} \ttt{iFunc}, \ttt{Interface}.
\itemc{Includes:} 
\ttt{<cmath>}
\ttt{<complex>}
\ttt{<cstdlib>}
\ttt{<iostream>}
\ttt{<fstream>}
\ttt{<map>}
\ttt{<string>}
\ttt{<utility>}
\ttt{<vector>}
\ttt{<cassert>}

\end{entry}

\drawbox{ExDiff::iFunc}\label{c:iFunc}
\begin{entry}
	\itemc{Purpose:} to initialize special class which contains a multidimensional interpolation function. Node points and values of the function are taken from \ttt{*.grid} (or by the use of some calculations) and \ttt{*.dat} files correspondingly. 
	\itemc{ }
	\itemc{Initialization example:}

\itemc{} \ttt{...}	
\itemc{} \ttt{std::vector<std::size\_t> dim;}
\itemc{} \ttt{dim = \{129,129,9,9\};}
\itemc{} \ttt{...}
\itemc{} \ttt{ExDiff::NDarray<double> dat\_(dim);}
\itemc{} \ttt{ExDiff::NDgrid<double>  grid\_(dim);}
\itemc{} \ttt{...}
\itemc{} \ttt{// load *.dat and *.grid}
\itemc{} \ttt{ExDiff::iFunc fungen(dat\_,grid\_);}	
\itemc{} \ttt{...}	

    \itemc{ }
	\itemc{Methods:}
	
	\itemc{ }
	\iteme{double Calc(std::vector<double>)}\label{m:iFunc.Calc} 
	\itemc{ } calculates the value of the interpolation function in the point defined by vector of variables. {\bf Example:}
	\itemc{ }
	\itemc{} \ttt{...}
	\itemc{ } \ttt{std::vector<double> p = \{1.2345,2.3456,3.4567,4.5678\};}
	\itemc{ } \ttt{double value = fungen.Calc(p);}
	\itemc{} \ttt{...}
	\itemc{ }
	
	\itemc{ }
	\iteme{int CheckPoint(std::vector<double>\& \_point)}\label{m:iFunc.CheckPoint} 
	\itemc{ } private method which returns \ttt{0} if vector \ttt{\_point} lies inside the interpolation range and \ttt{1} if it is outside the interpolation range.	
	\itemc{ }
	
	\iteme{NDgrid<double>\& GetGrid()}\label{m:iFunc.GetGrid} 
	\itemc{ } returns an object of the private class \ttt{iGrid}, which contains data from \ttt{*.grid} file. 
	\itemc{ }
	
	\iteme{NDarray<double>\& GetTab()}\label{m:iFunc.GetTab} 
	\itemc{ } returns an object of the private class \ttt{iTab}, which contains data from \ttt{*.dat} file. 
	\itemc{ }
	
	\iteme{std::vector<std::size\_t>\& GetDimensions()}\label{m:iFunc.GetDimensions} 
	\itemc{ } returns the vector of dimensions which is equal to \ttt{dim} in the initialization example.
	\itemc{ }		

	\iteme{std::vector<double>\& GetPoint()}\label{m:iFunc.GetPoint} 
	\itemc{ } returns the private vector \ttt{point} which contains the last point of calculations.
	\itemc{ }

	\iteme{std::vector<double>\& GetMinFunPoint()}\label{m:iFunc.GetMinFunPoint} 
	\itemc{ } returns private vector \ttt{minfunpoint} where the function is minimal.
	\itemc{ }	

	\iteme{std::vector<double>\& GetMaxFunPoint()}\label{m:iFunc.GetMaxFunPoint} 
	\itemc{ } returns private vector \ttt{maxfunpoint} where the function is maximal.
	\itemc{ }

	\iteme{double\& GetMinFun()}\label{m:iFunc.GetMinFun} 
	\itemc{ } returns private variable \ttt{minfunvalue} which is equal to the minimal value of the function.
	\itemc{ }
	
	\iteme{double\& GetMaxFun()}\label{m:iFunc.GetMaxFun} 
	\itemc{ } returns private variable \ttt{maxfunvalue} which is equal to the maximal value of the function.
	\itemc{ }	
	
	\iteme{void PrintVector(std::vector<size\_t>, std::size\_t)}
	\iteme{void PrintVector(std::vector<double>, std::size\_t)}\label{m:iFunc.PrintVector} 
	\itemc{ } prints \ttt{size\_t} or \ttt{double} vector
	\itemc{ }
	
	\iteme{void printC()}\label{m:iFunc.printC} 
	\itemc{ } prints the general information about the class object.
	\itemc{ }				
	
	\itemc{Variables:}
	\itemc{ }
	
	\iteme{std::vector<std::size\_t> dimensions :}\label{v:iFunc.dimensions}  
	private vector, which contains dimensions of the array (numbers of node points for each variable).
	\itemc{ }
	
	\iteme{std::vector<double> minvars :}\label{v:iFunc.minvars}  
	private vector, which contains minimal values of variables in the interpolation range.
	\itemc{ }	
	
	\iteme{std::vector<double> maxvars :}\label{v:iFunc.maxvars}  
	private vector, which contains maximal values of variables in the interpolation range.
	\itemc{ }	
	
	\iteme{std::vector<double> minfunpoint :}\label{v:iFunc.minfunpoint}  
	private vector, which contains point where the function is minimal.
	\itemc{ }
	
	\iteme{std::vector<double> maxfunpoint :}\label{v:iFunc.maxfunpoint}  
	private vector, which contains point where the function is maximal.
	\itemc{ }	
	
	\iteme{double minfunvalue :}\label{v:iFunc.minfunvalue}  
	private variable, which contains the minimal value of the function.
	\itemc{ }
	
	\iteme{double maxfunvalue :}\label{v:iFunc.maxfunvalue}  
	private variable, which contains the maximal value of the function.
	\itemc{ }				
	
	\itemc{Used by:} \ttt{Generator}, \ttt{Interface}.
	\itemc{Includes:} 
	\ttt{<cmath>}
	\ttt{<complex>}
	\ttt{<cstdlib>}
	\ttt{<iostream>}
	\ttt{<fstream>}
	\ttt{<map>}
	\ttt{<string>}
	\ttt{<utility>}
	\ttt{<vector>}
	\ttt{<cassert>}
	\ttt{"NDarray.h"}
	\ttt{"NDgrid.h"}
	\ttt{"I.h"}
	\ttt{"PI.h"}			
	
\end{entry}

%========================
\drawbox{ExDiff::Generator}\label{c:Generator}
\begin{entry}
	\itemc{Purpose:} to initialize special class which is applied to a multidimensional interpolation function, and generates a set of variables according to this function. \ttt{iFunc} class is used to create the input object.
	\itemc{ }
	\itemc{Initialization example:}
	
	\itemc{} \ttt{...}	
	\itemc{} \ttt{ExDiff::Generator gen(fungen);}
	\itemc{} \ttt{ExDiff::newrand();}
	\itemc{} \ttt{...}
    \itemc{}
    \itemc{} \ttt{fungen} is taken from the initialization example of the class \ttt{ExDiff::iFunc}. \ttt{ExDiff::newrand()} sets the random initial condition for new generation by the use of a system clock.
	
	\itemc{ }
	\itemc{Methods:}
	\itemc{ }

	\iteme{double RNUM()}\label{m:Generator.RNUM} 
	\itemc{ } returns random number between 0 and 1.
	\itemc{ }
	
	\iteme{void printC()}\label{m:Generator.printC} 
	\itemc{ } prints the general information about the object of the class.
	\itemc{ }				

	\iteme{ExDiff::iFunc CalcMinus(const ExDiff::iFunc\&)}\label{m:Generator.CalcMinus} 
	\itemc{ } calculates new \ttt{iFunc} object of $N-1$ variables by resummation in the last variable from the vector of 
	variables, where $N$ is the number of variables of the input object.
	\itemc{ }
	
	\iteme{std::vector<ExDiff::iFunc> CalcFUNG(const ExDiff::iFunc\&)}\label{m:Generator.CalcFUNG} 
	\itemc{ } calculates a full set of interpolation functions of lower dimensions, including also the total integral of the input interpolation function, and outputs results to private vector \ttt{FUNG} and the variable \ttt{FUNGTOT}.
	\itemc{ }	
	
	\iteme{double\& GetFUNGTOT()}\label{m:Generator.GetFUNGTOT} 
	\itemc{ } gets the total cross-section from the private variable \ttt{FUNGTOT}.
	\itemc{ }	
	
	\iteme{double GenVar(const std::vector<double>\& x\_,}
	\iteme{const std::vector<double>\& f\_, const double\& R\_)}\label{m:Generator.GenVar} 
	\itemc{ } generates a variable from one-dimensional interpolation function, represented by the \ttt{std::vector} array \ttt{f\_}. \ttt{x\_} contains node points, \ttt{f\_} contains values of the function in these node points, \ttt{R\_} is the input random number. 
	\itemc{ }

	\iteme{double EGenVarFast1(const std::vector<double>\& x\_,}
	\iteme{const std::vector<double>\& f\_, const std::vector<double>\& F\_,}
	\iteme{const double\& R\_)}\label{m:Generator.GenVarFast1} 
	\itemc{ } generates a variable from one-dimensional interpolation function, represented by the \ttt{std::vector} array \ttt{f\_}. \ttt{x\_} contains node points, \ttt{f\_} contains values of the function in these node points, \ttt{F\_} contains integrals of the function, \ttt{R\_} is the input random number.
	\itemc{ }	
	
	\iteme{double IntVec(const std::vector<double>\& x\_, const std::vector<double>\& f\_)}\label{m:Generator.IntVec} 
	\itemc{ } calculates the total integral of one-dimensional interpolation function. \ttt{x\_} contains node points, \ttt{f\_} contains values of the function in these node points.	
	\itemc{ }
	
	\iteme{std::vector<double> Generate()}\label{m:Generator.Generate} 
	\itemc{ } generates the output set (\ttt{std::vector}) of variables.
	\itemc{ }	
	
	\itemc{Variables:}
	\itemc{ }
	
	\iteme{ExDiff::iFunc FUN :}\label{v:Generator.FUN}  
	private variable which contains an input object of class \ttt{iFunc} (initial multidimensional interpolation function).
	\itemc{ }
	
	\iteme{std::vector<ExDiff::iFunc> FUNG :}\label{v:Generator.FUNG}  
	private vector which contains an input object of class \ttt{iFunc} (initial multidimensional interpolation function) and all the functions of lower dimensions.
	\itemc{ }			
	
	\iteme{double FUNGTOT :}\label{v:Generator.FUNGTOT}  
	private variable which contains the total cross-section.
	\itemc{ }	
	
	\itemc{Used by:} \ttt{Interface}.
	\itemc{Includes:} 
	\ttt{<cmath>}
	\ttt{<complex>}
	\ttt{<cstdlib>}
	\ttt{<iostream>}
	\ttt{<fstream>}
	\ttt{<map>}
	\ttt{<string>}
	\ttt{<utility>}
	\ttt{<vector>}
	\ttt{<cassert>}
	\ttt{<time.h>}
	\ttt{<ctime>} 	
	
	\ttt{"NDarray.h"}
	\ttt{"NDgrid.h"}
	\ttt{"iFunc.h"}
	\ttt{"newrand.h"}	
	\ttt{"I.h"}
	\ttt{"PI.h"}			
	
\end{entry}
%========================

%========================
\drawbox{ExDiff::Constants}\label{c:Constants}
\begin{entry}
	\itemc{Purpose:} to initialize special class which contains fundamental constants: couplings, masses, spins and IDs of particles. 
	\itemc{ }
	\itemc{Initialization example:}
	
	\itemc{} \ttt{...}	
	\itemc{} \ttt{ExDiff::Constants c;}
	\itemc{} \ttt{...}
	\itemc{}
	
	\itemc{Methods:}
	\itemc{}
	
	\iteme{void Print()}\label{m:Constants.Print} 
	\itemc{ } prints all the constants.
	\itemc{ }				
	
	\iteme{double \_mass(int ID\_)}\label{m:Constants.mass} 
	\itemc{ } returns the mass of a particle by the use of its ID number.
	\itemc{ }
	
	\iteme{int \_dspin(int ID\_)}\label{m:Constants.dspin} 
	\itemc{ } returns $2s+1$, where $s$ is the particle spin.
	\itemc{ }	
	
	\itemc{Variables:}
	\itemc{ }
	
	\iteme{const double m\_p, m\_n, m\_pi0, m\_pi, m\_H, m\_Gra, m\_R, m\_Glu, m\_Z, m\_W :}\label{v:Constants.mass}  
	public variables which contain masses of proton, neutron, $\pi_0$, $\pi^{\pm}$, Higgs boson, graviton, glueball, Z and W bosons.
	\itemc{ }

	\iteme{const double  alpha\_EM, Lam\_QCD :}\label{v:Constants.couplings}  
	public variables which contain electromagnetic $\alpha_e$ and QCD $\alpha_s$ couplings.
	\itemc{ }	

	\iteme{const double VEV, M\_Pl :}\label{v:Constants.ewpars}  
	public variables which contain vacuum expectation value $246$~GeV and the Plank mass.
	\itemc{ }
	
	\iteme{const int IDproton, ID1710, ID1950, IDpi0, IDpiplus, IDpiminus, IDdef :}\label{v:Constants.IDs}  
	public variables which contain PDG (or PYTHIA's) IDs of particles.
	\itemc{ }		
	
	\itemc{Used by:} \ttt{Interface}, \ttt{Event}, \ttt{KinCM}.
	\itemc{Includes:} 
	\ttt{<cmath>}
	\ttt{<complex>}
	\ttt{<cstdlib>}
	\ttt{<iostream>}
	\ttt{<fstream>}
	\ttt{<map>}
	\ttt{<string>}
	\ttt{<utility>}
	\ttt{<vector>}
	\ttt{<cassert>}	
	
	\ttt{"I.h"}
	\ttt{"PI.h"}			
	
\end{entry}
%========================

%========================
\drawbox{ExDiff::Particle}\label{c:Particle}
\begin{entry}
	\itemc{Purpose:} to initialize special class which contains four-momentum, mass, spin, color, ID of a relativistic particle. 
	\itemc{ }
	\itemc{Initialization example:}

	\itemc{} \ttt{...}	
	\itemc{} \ttt{ExDiff::Constants c;}  
	\itemc{} \ttt{std::vector<double> v = \{c.m\_p,0.0,0.0,0.0\};}
	\itemc{} \ttt{ExDiff::Particle pa(v,c.m\_p,2212,2,0,0);}
	\itemc{} \ttt{...}	
	
	\itemc{Methods:}
	\itemc{}
	
	\iteme{void Print()}\label{m:Particle.Print} 
	\itemc{ } prints all the information about the particle.
	\itemc{ }				
	
	\iteme{double Rapidity(std::vector<double>\& p\_)}\label{m:Particle.Rapidity} 
	\itemc{ } calculates particle rapidity. \ttt{p\_} is the input four-momentum. 
	\itemc{ }	
	
	\iteme{double Pseudorapidity(std::vector<double>\& p\_)}\label{m:Particle.Pseudorapidity} 
	\itemc{ } calculates particle pseudorapidity. \ttt{p\_} is the input four-momentum.
	\itemc{ }
	
	\iteme{double Theta(std::vector<double>\& p\_)}\label{m:Particle.Theta} 
	\itemc{ } calculates particle polar angle. \ttt{p\_} is the input four-momentum.
	\itemc{ }
	
	\iteme{double Phi(std::vector<double>\& p\_)}\label{m:Particle.Phi} 
	\itemc{ } calculates particle azimuthal angle. \ttt{p\_} is the input four-momentum.
	\itemc{ }			
	
	\iteme{double Lmult(std::vector<double>\& p1\_, std::vector<double>\& p2\_)}\label{m:Particle.Lmult} 
	\itemc{ } returns scalar product of \ttt{p1\_} and \ttt{p2\_} four momenta. 
	\itemc{ }
	
	\iteme{std::vector<double> Lminus(std::vector<double>\& p1\_, std::vector<double>\& p2\_)}\label{m:Particle.Lminus} 
	\itemc{ } returns four momenta which is equal to \ttt{p1\_} minus \ttt{p2\_}. 
	\itemc{ }	

	\iteme{std::vector<double> Lplus(std::vector<double>\& p1\_, std::vector<double>\& p2\_)}\label{m:Particle.Lplus} 
	\itemc{ } returns four momenta which is equal to \ttt{p1\_} plus \ttt{p2\_}. 
	\itemc{ }	

	\iteme{void Lprint(std::vector<double>\& vec\_)}\label{m:Particle.Lprint} 
	\itemc{ } prints the four-momentum \ttt{vec\_}.
	\itemc{ }

	\iteme{int OnShell()}\label{m:Particle.OnShell} 
	\itemc{ } returns on-shell status of the particle:
	\begin{subentry}
		\iteme{0  :} particle is on-shell, $p^2=m^2$
		\iteme{1  :} $0<p^2<m^2$
		\iteme{-1 :} $-m^2<p^2<0$ 
		\iteme{2  :} $p^2>m^2$
		\iteme{-2 :} $p^2<-m^2$
		\iteme{3  :} $m=0$
		\iteme{4  :} $m<0$ (error value)
	\end{subentry}		
	\itemc{ }

	\iteme{void Reset(const std::vector<double>\& p\_, const double\& m\_,}
	\iteme{const int\& ID\_, const int\& DSpin\_,}
	\iteme{const int\& Colour\_, const int\& Anticolour\_)}\label{m:Particle.Reset} 
	\itemc{ } resets parameters of a particle (all the private variables).
	\itemc{ }

	\iteme{void ResetP(const std::vector<double>\& p\_)}\label{m:Particle.ResetP} 
	\itemc{ } resets only four-momentum of the particle.
	\itemc{ }
	
	\iteme{const std::vector<double>\& \_p()}\label{m:Particle.p} 
	\itemc{ } returns private variable which contains four-momentum of the particle.
	\itemc{ }	
	
	\iteme{double\& \_px()}
    \iteme{double\& \_py()}
    \iteme{double\& \_pz()}
    \iteme{double\& \_E()}
    \iteme{double\& \_m()}
    \iteme{double\& \_pp()}
    \iteme{double\& \_pT()}
    \iteme{double\& \_p3D()}
    \iteme{int\& \_DSpin()}
    \iteme{int\& \_ID()}
    \iteme{int\& \_Colour()}
    \iteme{int\& \_Anticolour()}	
	\label{m:Particle.vars} 
	\itemc{ } return private variables which contain $p_x$, $p_y$, $p_z$, $E$, $p^2$, $p_T=\sqrt{p_x^2+p_y^2}$, $p_{3D}=\sqrt{p_x^2+p_y^2+p_z^2}$, $2s+1$ ($s$ is the spin), ID number, colour and anticolour of the particle.
	\itemc{ }	
				
	\itemc{Variables:}
	\itemc{ }
	
	\iteme{std::vector<double> p :}\label{v:Particle.p}   
	private vector which contains four-momentum of the particle ${E,p_x,p_y,p_z}$.
	\itemc{ }
	
	\iteme{double E, px, py, pz, m, pp, pT, p3D  :}\label{v:Particle.vars}  
	private variables which contain $E$, $p_x$, $p_y$, $p_z$, $m$, $p^2$, $p_T=\sqrt{p_x^2+p_y^2}$, $p_{3D}=\sqrt{p_x^2+p_y^2+p_z^2}$.
	\itemc{ }	

	\iteme{int ID, DSpin, Colour, Anticolour  :}\label{v:Particle.intvars}  
	private variables which contain ID, $2s+1$, colour and anticolour of the particle.
	\itemc{ }			
	
	\itemc{Used by:} \ttt{Interface}, \ttt{Event}, \ttt{KinCM}.
	\itemc{Includes:} 
	\ttt{<cmath>}
	\ttt{<complex>}
	\ttt{<cstdlib>}
	\ttt{<iostream>}
	\ttt{<fstream>}
	\ttt{<map>}
	\ttt{<string>}
	\ttt{<utility>}
	\ttt{<vector>}
	\ttt{<cassert>}	
	
	\ttt{"I.h"}
	\ttt{"PI.h"}			
	
\end{entry}
%========================

%========================
\drawbox{ExDiff::Event}\label{c:Event}
\begin{entry}
	\itemc{Purpose:} to initialize special class which contains all the particles of an event. 
	\itemc{ }
	\itemc{Initialization example:}
	
	\itemc{} \ttt{...}	
	\itemc{} \ttt{ExDiff::Event ev(particles);}
	\itemc{} \ttt{...}
	\itemc{}
	\itemc{} where \ttt{particles} is of the \ttt{std::vector<Particle>} type.
	
	\itemc{Methods:}
	\itemc{}
	
	\iteme{void Print(int format)}\label{m:Event.Print} 
	\itemc{ } prints parameters of particles in different form.
	\begin{subentry}
		\iteme{format=0 :} all the parameters of every particle.
		\iteme{format=1 :} mass, on-shell status and four-momentum of every particle.
		\iteme{format=2 :} four-momenta of particles.
		\iteme{format=3 :} four-momenta of final particles.
	\end{subentry}		
	\itemc{ }				
	
	\iteme{std::vector<Particle>\& \_particles()}\label{m:Event.particles} 
	\itemc{ } returns private vector \ttt{particles}.
	\itemc{ }
	
	\iteme{void AddToFile(int format\_, std::string filename\_)}\label{m:Event.AddToFile} 
	\itemc{ } add an event to file \ttt{*.exdiff} in special format.
	\begin{subentry}
		\iteme{format\_=0 :} intrinsic \ExD\ format.
	\end{subentry}		
	\itemc{ }
	
	\iteme{std::vector<int> ExDiff::Event::TakeFromFile(int number\_,}
	\iteme{int format\_, std::string filename\_)}\label{m:Event.TakeFromFile} 
	\itemc{ } loads an event number \ttt{number\_} from the file \ttt{filename\_} in special format.
	\begin{subentry}
		\iteme{format\_=0 :} intrinsic \ExD\ format.
	\end{subentry}		
	\itemc{ } returns \ttt{int} vector, which contains numbers of 
	\begin{subentry}
		\iteme{[0] :} particles in the event,
		\iteme{[1] :} events in the file,
		\iteme{[2] :} strings in the file.				
	\end{subentry}		
	\itemc{ }	
	
	\iteme{void ExDiff::Event::TakeEvent(std::vector<int> Nfile\_,}
	\iteme{int number\_, int format\_, std::string filename\_)}\label{m:Event.TakeEvent} 
	\itemc{ } loads an event number \ttt{number\_} from the file \ttt{filename\_} in special format.
	\begin{subentry}
		\iteme{format\_=0 :} intrinsic \ExD\ format.
	\end{subentry}		
	\itemc{ } The file is \textbf{the same} as in the previouse method. Vector \ttt{Nfile\_} is obtained by the use
	of \ttt{TakeFromFile}: 	
	\itemc{} \ttt{...}	
	\itemc{} \ttt{std::vector<int> Nfile = ev.TakeFromFile(ev\_number,format,filename);}
	\itemc{} \ttt{ev.TakeEvent(Nfile,ev\_number,format,filename);}	
%	\itemc{} \ttt{...}
	\itemc{}					

	\iteme{int CheckSum()}\label{m:Event.CheckSum} 
	\itemc{ } returns $0$, if momentum conservation law is carried out, $1$ in other cases.
	\itemc{ }		
	
	\itemc{Variables:}
	\itemc{ }
	
	\iteme{std::vector<ExDiff::Particle> particles :}\label{v:Event.particles}  
	private vector which contains all the particles.
	\itemc{ }		
	
	\itemc{Used by:} \ttt{Interface}.
	\itemc{Includes:} 
	\ttt{<cmath>}
	\ttt{<complex>}
	\ttt{<cstdlib>}
	\ttt{<iostream>}
	\ttt{<fstream>}
	\ttt{<map>}
	\ttt{<string>}
	\ttt{<utility>}
	\ttt{<vector>}
	\ttt{<cassert>}
	\ttt{<iomanip>}
	\ttt{<limits>}		
	
	\ttt{"I.h"}
	\ttt{"PI.h"}
	\ttt{"Particle.h"}
	\ttt{"Constants.h"}				
	
\end{entry}
%========================

%========================
\drawbox{ExDiff::KinCM}\label{c:KinCM}
\begin{entry}
	\itemc{Purpose:} to initialize special class which contains all kinematical variables for different processes. 
	\itemc{ }
	\itemc{Initialization example:}
	
	\itemc{} \ttt{...}	
	\itemc{} \ttt{ExDiff::KinCM kinematics(sqs\_,0,genvec,masses\_\_,ids\_\_);}
	\itemc{} \ttt{...}
	\itemc{}
	\itemc{} where \ttt{sqs\_} is the central mass frame collision energy, $0$ defines the set of variables, \ttt{genvec} contains generated variables, \ttt{masses\_\_} and \ttt{ids\_\_} contain all the {\bf final(!)} particles masses and IDs. Other way to initialize an object of the class is
	\itemc{} \ttt{...}	
	\itemc{} \ttt{ExDiff::KinCM kinematics(sqs\_,particles\_);}
	\itemc{} \ttt{...}
	\itemc{}
	\itemc{} where \ttt{particles\_} is the vector of type \ttt{ExDiff::Particle} which contains all the particles of a process.
	\itemc{}
	
	\itemc{Methods:}
	\itemc{}
	
	\iteme{void Print()}\label{m:KinCM.Print} 
	\itemc{ } prints all the information on kinematical variables of the process.	
	\itemc{ }	

	\iteme{std::vector<Particle> EvParticles()}\label{m:KinCM.EvParticles} 
	\itemc{ } returns the vector of all the particles including initial protons 
	(to use it in the \ttt{ExDiff::Event} class objects).	
	\itemc{ }
	
	\iteme{void ResetVars(const std::vector<double>\& variables\_)}\label{m:KinCM.ResetVars} 
	\itemc{ } sets new values only to kinematical variables of the process without changing 
	masses, ids, spins and so on. Also it verifies conservation laws.	
	\itemc{ }
	
	\iteme{int Phys()}\label{m:KinCM.Phys} 
	\itemc{ } checks out physical region conditions for the process (conservation laws, positivity of 
	particles energies, all particles are on-shell) and returns
	\begin{subentry}
		\iteme{0 :} everything is correct
		\iteme{1 :} some of energies are negative
		\iteme{2 :} some particles are off-shell
		\iteme{>2 :} other errors
	\end{subentry}	
	\itemc{ }		
	
	\iteme{std::vector<double> Transform\_CMpp\_to\_CMdd(std::vector<double>\& k\_)}\label{m:KinCM.TransformPD} 
	\itemc{ } makes transformation of the Lorentz vector \ttt{k\_} calculated in the C.M. frame of two colliding protons to
	the frame where $\vec{\Delta}_1+\vec{\Delta}_2=0$ and ${\Delta_1}_0+{\Delta_2}_0=\sqrt{(\Delta_1+\Delta_2)^2}$ 
	(the rest frame of a central resonance or system of particles).	All the vectors are defined in~(\ref{kin:momenta22}).		
	\itemc{ }	
	
	\iteme{std::vector<double> Transform\_CMdd\_to\_CMpp(std::vector<double>\& k\_)}\label{m:KinCM.TransformDP} 
	\itemc{ } makes transformation of the Lorentz vector \ttt{k\_} calculated in the 
	reference frame where $\vec{\Delta}_1+\vec{\Delta}_2=0$ and 
	${\Delta_1}_0+{\Delta_2}_0=\sqrt{(\Delta_1+\Delta_2)^2}$ 
	(the rest frame of a central resonance or system of particles) to the C.M. frame of two colliding protons. 
	All the vectors are defined in~(\ref{kin:momenta22}).			
	\itemc{ }					

	\iteme{std::vector<double> Transform\_CMpp\_to\_CMdd\_(std::vector<double>\& k\_,}
	\iteme{std::vector<double>\& Delta\_1\_, std::vector<double>\& Delta\_2\_)}\label{m:KinCM.TransformPD1} 
	\itemc{ } makes transformation of the Lorentz vector \ttt{k\_} calculated in the C.M. frame of two colliding protons to
	the frame where $\vec{\Delta}_1+\vec{\Delta}_2=0$ and ${\Delta_1}_0+{\Delta_2}_0=\sqrt{(\Delta_1+\Delta_2)^2}$ 
	(the rest frame of a central resonance or system of particles). \ttt{Delta\_1\_} and \ttt{Delta\_2\_} contain 
	vectors $\Delta_{1,2}$ calculated in the C.M. frame of two colliding protons. All the vectors are defined in~(\ref{kin:momenta22}).			
	\itemc{ }	
	
	\iteme{std::vector<double> Transform\_CMdd\_to\_CMpp\_(std::vector<double>\& k\_)}
	\iteme{std::vector<double>\& Delta\_1\_, std::vector<double>\& Delta\_2\_)}\label{m:KinCM.TransformDP1} 
	\itemc{ } makes transformation of the Lorentz vector \ttt{k\_} calculated in the 
	reference frame where $\vec{\Delta}_1+\vec{\Delta}_2=0$ and 
	${\Delta_1}_0+{\Delta_2}_0=\sqrt{(\Delta_1+\Delta_2)^2}$ 
	(the rest frame of a central resonance or system of particles) to the C.M. frame of two colliding protons. \ttt{Delta\_1\_} 
	and \ttt{Delta\_2\_} contain vectors $\Delta_{1,2}$ calculated in the C.M. frame of two colliding protons. All the vectors are defined in~(\ref{kin:momenta22}).			
	\itemc{ }		

	\iteme{void VtoP2(std::size\_t\& \_ID, std::vector<double>\& \_masses,}
	\iteme{std::vector<int>\& \_IDs)}\label{m:KinCM.VtoP2} 
	\itemc{ } sets all the final particles momenta from the set of variables for the process $2\to 2$. \ttt{\_masses} and \ttt{\_IDs} are vectors of masses and ids of two final particles. \ttt{\_ID} identifies the set of kinematical variables for the process:
	\begin{subentry}
		\iteme{\ttt{\_ID}=0 :} $|t|$, $\phi$;
		\iteme{\ttt{\_ID}=1 :} $\theta$ (polar scattering angle of the final hadron), $\phi$ (azimuthal angle of the final hadron);
		\iteme{\ttt{\_ID}=2 :} $y_h$ (rapidity of the final hadron), $\phi$;
		\iteme{\ttt{\_ID}=3 :} $|\vec{\Delta}|$ (transverse momentum of the final hadron), $\phi$;
	\end{subentry}		 	
	\itemc{ }
	
	\iteme{void VtoP3(std::size\_t\& \_ID, std::vector<double>\& \_masses,}
	\iteme{std::vector<int>\& \_IDs)}\label{m:KinCM.VtoP3} 
	\itemc{ } similar to previous one, but it sets all the final particles momenta from the set of variables for the process $2\to 3$. \ttt{\_ID} identifies the set of kinematical variables for the process:
	\begin{subentry}
		\iteme{\ttt{\_ID}=0 :} $|\vec{\Delta}_{1T}|$, $|\vec{\Delta}_{2T}|$, $\phi_0$ (azimuthal angle between final protons), $\ln\xi_1$, $\phi_1$\\ (azimuthal angle of the first final hadron);
		\iteme{\ttt{\_ID}=1 :} $|\vec{\Delta}_{1T}|$, $|\vec{\Delta}_{2T}|$, $\phi_0$, $y_c$ (rapidity of a central particle), $\phi_1$;
	\end{subentry}		 	
	\itemc{ }	
	
	\iteme{void VtoP4(std::size\_t\& \_ID, std::vector<double>\& \_masses,}
	\iteme{std::vector<int>\& \_IDs)}\label{m:KinCM.VtoP4} 
	\itemc{ } similar to previous one, but it sets all the final particles momenta from the set of variables for the process $2\to 4$. \ttt{\_ID} identifies the set of kinematical variables for the process:
	\begin{subentry}
		\iteme{\ttt{\_ID}=0 :} $|\vec{\Delta}_{1T}|$, $|\vec{\Delta}_{2T}|$, $\phi_0$, $\phi_1$, $M$ (mass of a central system), $y$ (rapidity of the central system), $\eta_j$ (pseudorapidity of the first particle in the central system in its rest frame, or, in other words, in the C.M. frame of two central particles. For example, consider $\pi^+\pi^-$ central production. In this case $\eta_j\equiv\eta_{\pi^+}$ or $\eta_j\equiv\eta_{\pi^-}$ in the $\pi^+\pi^-$ C.M. frame. It is equal to $\eta$ in~(\ref{kin:momenta23})), $\phi_j$ (azimuthal angle of the particle, which we choose to fix $\eta_j$, in the same reference frame);
		\iteme{\ttt{\_ID}=1 :} $|\vec{\Delta}_{1T}|$, $|\vec{\Delta}_{2T}|$, $\phi_0$, $\phi_1$, $\xi_1$, $\xi_2$ (see~(\ref{kin:momenta22})), $\eta_j$, $\phi_j$;
	\end{subentry}		 	
	\itemc{ }		

	\iteme{void PtoV2()}\label{m:KinCM.PtoV2} 
	\itemc{ } calculates all the variables for a $2\to 2$ process in the second initialization case (see above), when we use momenta as input. 	
	\itemc{ }	

	\iteme{void PtoV3()}\label{m:KinCM.PtoV3} 
	\itemc{ } calculates all the variables for a $2\to 3$ process in the second initialization case (see above), when we use momenta as input. 	
	\itemc{ }
	
	\iteme{void PtoV4()}\label{m:KinCM.PtoV4} 
	\itemc{ } calculates all the variables for a $2\to 4$ process in the second initialization case (see above), when we use momenta as input. 	
	\itemc{ }		
	
	\iteme{X \_Y()}\label{m:KinCM.GetY} 
	\itemc{ } give the access to the private variable \ttt{Y} of the type (class) \ttt{X} (see below the list of private variables). For example \ttt{std::vector<Particle>\& \_particles()} returns private vector \ttt{particles}, \ttt{double\& \_Eta\_j()} returns \ttt{Eta\_j} variable.	
	\itemc{ }	
	
	\itemc{Variables:}
	\itemc{ }
	
	\iteme{ExDiff::Particle hadron1, hadron2 :}\label{v:KinCM.hadrons}  
	private objects which contain information on initial hadrons.
	\itemc{ }		
	
	\iteme{std::vector<Particle> particles :}\label{v:KinCM.particles}  
	private object which contains information on final particles.
	\itemc{ }	
	
	\iteme{std::vector<double> variables :}\label{v:KinCM.variables}  
	private vector which contains values of variables used in the input set.
	\itemc{ }

	\iteme{double sqs :}\label{v:KinCM.sqs}  
	private variable which contains $\sqrt{s}$ (collision energy).
	\itemc{ }		
	
	\iteme{double t, phi, DeltaT, Delta3D, theta, y}
	\iteme{std::vector<double> Delta :}\label{v:KinCM.vars22}  
	private variables which contain $|t|$, $\phi$, $|\vec{\Delta}|=\sqrt{\Delta_x^2+\Delta_y^2}$, $\sqrt{\Delta_x^2+\Delta_y^2+\Delta_z^2}$, $\theta$, $y$, $\Delta$ of a $2\to 2$ process (see~(\ref{kin1c})).
	\itemc{ }

\iteme{double t\_1, t\_2, phi\_12, phi\_1, xi\_1, xi\_2, y\_c, M\_T, M\_c,} 
\iteme{DeltaT\_1, DeltaT\_2, Delta3D\_1, Delta3D\_2}
\iteme{std::vector<double> Delta\_1, Delta\_2 :}\label{v:KinCM.vars23}  
	private variables which contain $|t_1|$, $|t_2|$, $\phi_0$, $\phi_1$, $\xi_1$, $\xi_2$, $y$, $M_{\perp}$, $M$, 
	$|\vec{\Delta}_1|$, $|\vec{\Delta}_2|$, $\sqrt{\Delta_{1x}^2+\Delta_{1y}^2+\Delta_{1z}^2}$,  $\sqrt{\Delta_{2x}^2+\Delta_{2y}^2+\Delta_{2z}^2}$, $\Delta_1$, $\Delta_2$ (see~(\ref{kin:momenta22}),(\ref{kin:relations})).
	\itemc{ }
	
	\iteme{double t\_hat, s\_hat, Eta\_j, Theta\_j, Phi\_j :}\label{v:KinCM.vars24}  
	additional private variables which contain $\hat{t}_a$, $\hat{s}=M^2$, $\eta_j=\eta=-\ln\tan\frac{\theta_j}{2}$, $\theta_j$, $\phi_j$ (see~(\ref{kin:momenta23}),(\ref{kin:relations2})).
	\itemc{ }
	
	\iteme{int flag :}\label{v:KinCM.flag}  
	private variable which contains $0$ if everything is correct in the kinematics and nonzero values in other cases.
	\itemc{ }
	
	\iteme{std::size\_t ID}
	\iteme{std::vector<double> masses}
	\iteme{std::vector<int> IDs}		
	\label{v:KinCM.varsinput}  
	private variables (vectors) which contain information on the set of input variables, masses and ids of final particles.
	\itemc{ }						
	
	\itemc{Used by:} \ttt{Interface}.
	\itemc{Includes:} 
	\ttt{<cmath>}
	\ttt{<complex>}
	\ttt{<cstdlib>}
	\ttt{<iostream>}
	\ttt{<fstream>}
	\ttt{<map>}
	\ttt{<string>}
	\ttt{<utility>}
	\ttt{<vector>}
	\ttt{<stdio.h>}
	\ttt{<cassert>}	
	
	\ttt{"I.h"}
	\ttt{"PI.h"}
	\ttt{"Particle.h"}
	\ttt{"Constants.h"}				
	
\end{entry}
%========================

%========================
\drawbox{ExDiff::Interface}\label{c:Interface}
\begin{entry}
	\itemc{Purpose:} to initialize special class which provides convenient input/output. 
	\itemc{ }
	\itemc{Initialization example:}
	
	\itemc{} \ttt{...}	
	\itemc{} \ttt{ExDiff::Interface iogen(datacard);}
	\itemc{} \ttt{...}
	\itemc{}
	\itemc{} where \ttt{datacard} is the name of a configuration (\ttt{*.card}) file.
	\itemc{}
	
	\itemc{Methods:}
	\itemc{}
	
	\iteme{void GeneratorInfo()}\label{m:Interface.GeneratorInfo} 
	\itemc{ } prints the general information on the generator (authors, version and so on).	
	\itemc{ }		
	
	\iteme{void PrintPars()}\label{m:Interface.PrintPars} 
	\itemc{ } prints the general information on the process (authors, process, C.M. energy, version, number 
	of output events, input/output format).	
	\itemc{ }	
	
	\iteme{void GetCard(const char *\_filename)}\label{m:Interface.GetCard} 
	\itemc{ } reads parameters from the configuration file.		
	\itemc{ }	
	
	\iteme{void GenerateFile()}\label{m:Interface.GenerateFile} 
	\itemc{ } creates final output file according to input data.		
	\itemc{ }
	
	\iteme{std::string ResultFileName()}\label{m:Interface.ResultFileName} 
	\itemc{ } creates output filename which is used by default.		
	\itemc{ }	
	
	\iteme{std::string DatFileName()()}\label{m:Interface.DatFileName} 
	\itemc{ } creates input \ttt{*.dat} filename .		
	\itemc{ }
	
	\iteme{std::string GridFileName()}\label{m:Interface.GridFileName} 
	\itemc{ } creates input \ttt{*.grid} filename.		
	\itemc{ }					
			
	\iteme{int\& \_X()}\label{m:Interface.X} 
	\itemc{ } returns private variable \ttt{X}.	For example, \ttt{\_Nevents} returns the private 
	variable \ttt{Nevents} (see below).	
	\itemc{ }		
		
	\itemc{Variables:}
	\itemc{ }
	
	\iteme{int IDauthors, IDprocess, IDenergy, version, Nevents, kinformat, outformat:}\label{v:Interface.vars}  
	private variables which define authors, process, C.M. energy, version of the data files, number of 
	events, set of input variables and an output format.
	\itemc{ }		
	
	\itemc{Used by:} \ttt{main}.
	\itemc{Includes:} 
	\ttt{<fstream>}
	\ttt{<cassert>}
	\ttt{<cmath>}
	\ttt{<complex>}
	\ttt{<cstdlib>}
	\ttt{<iostream>}
	\ttt{<map>}
	\ttt{<string>}
	\ttt{<utility>}
	\ttt{<vector>}
	\ttt{<stdio.h>}
	\ttt{<time.h>}
	\ttt{<iomanip>}
	\ttt{<limits>}
	\ttt{<string>}
	\ttt{<ctime>}	
	\ttt{"newrand.h"}
	\ttt{"NDarray.h"}
	\ttt{"NDgrid.h"}
	\ttt{"iFunc.h"}
	\ttt{"Generator.h"}
	\ttt{"Particle.h"}
	\ttt{"Constants.h"}
	\ttt{"Kinematics.h"}
	\ttt{"Event.h"}			
	
\end{entry}
%========================

\newpage
\section{Program Installation}
\label{s:install}
 
Some materials related to the \ExD\ physics and generator can be found 
on the web page\\[2mm]
\drawbox{\ttt{https://exdiff.hepforge.org/}}\\
To get the code of the generator one should download the file \\[2mm]  
\drawbox{\ttt{https://exdiff.hepforge.org/code/ExDiff1.0.zip}}\\
The program is written essentially entirely in standard c++,
and should run on any platform with such a compiler. 

The following installation procedure is suggested for the Linux users,
it was tested with CERN SLC7. 

\begin{enumerate}
\item Copy files to your folder
\item Change the card file in the folder \ttt{config} or use sample files.
\item Go to the folder with Makefile
\item \ttt{\$ make}
\item \ttt{\$ ExDiff <cardfile name> [<output file name>]}
\item \ttt{\$ cd output}
\item see output file in the folder \ttt{output}
\end{enumerate}

\noindent
In the main folder you can see some files:\vskip 1mm
\begin{entry}
\iteme{readme.txt} contains brief description of the generator;
\iteme{main.cpp} is the main function;
\iteme{Makefile} defines compilation instructions;
\iteme{main\_with\_analizer.cpp} is the main function with analyzer examples;
\end{entry}

\newpage
\section{Getting Started with the Simple Example}
\label{s:start}

The Simple Example could look as following:

\begin{verbatim} 
/* main.cpp */
#include <fstream>
#include <cassert>
#include <cmath>
#include <complex>
#include <cstdlib>
#include <iostream>
#include <map>
#include <string>
#include <utility>
#include <vector>
#include <stdio.h>
#include <time.h>
#include <iomanip>
#include <limits>
#include <string>
#include <ctime>
#include "inc/I.h"
#include "inc/PI.h"
#include "inc/newrand.h"
#include "inc/NDarray.h"
#include "inc/NDgrid.h"
#include "inc/iFunc.h"
#include "inc/Generator.h"
#include "inc/Particle.h"
#include "inc/Constants.h"
#include "inc/Kinematics.h"
#include "inc/Event.h"
#include "inc/Interface.h"

using namespace ExDiff;

int main(int argc, const char** argv){

// final construction
// datacard is the argument of the main file (filename)
const char* datacard = argv[1]; 
// reading input data card
// create interface   
Interface iogen(datacard);
iogen.GeneratorInfo();
iogen.PrintPars();
// output using interface     
iogen.GenerateFile();

// Additional procedures

return 0;
}            
\end{verbatim} 

First, we set the object of the \ttt{ExDiff::Interface} class, then print the general information, and finally
generate an output file. If you need another type of output, you can make some modifications in the \ttt{ExDiff::Interface::GenerateFile} method (see subsection~\ref{ss:developers}).

\subsection{Analyzer}
\label{ss:analyzer}

Here you can see example of the simple analyzer. It can be inserted into the previous example instead
of the phrase \ttt{// Additional procedures}.

\begin{verbatim} 
// time for calculations start	//////////////////////////												
   printf("analizer clock start = %d\n", std::clock());
   start = std::clock();
/////////////////////////////////////////////////////////

   Event ev;
// input file
   std::string filename="output/A1M2E0_2F0.exdiff";
// output file
   std::string ofilename="output/t_A1M2E0_2F0.res";

   int i,j,n,Nbins;
   Nbins = 128; // number of bins in t variable
// t-bins 
   std::vector<double> tbin, result, tgrid;
   tbin.clear();    
   tgrid.clear();
   double taux=0.0;
   for(i = 0; i < Nbins; i++){
    taux=(2.0*i+1.0)*0.5/Nbins; 
    tbin.push_back(taux);
    taux=i*1.0/Nbins;
    tgrid.push_back(taux); 
    result.push_back(0.0);		        
   }      
\end{verbatim}
\begin{verbatim}
   std::cout<< "Begin to read events" <<std::endl;
// events reading and analysis
    
    int format = 0; // format of the input file
    double t; // |t| variable
    double x = 0.0; // auxiliary variable

// defines number of events in the file    
    std::vector<int> Nfile = ev.TakeFromFile(10,format,filename); 
    int Nevents = Nfile[1];

// prints parameters    
    std::cout<< "Number of events = " << Nevents 
    << " Number of particles = " << Nfile[0] 
    << " Number of strings = " << Nfile[2] <<std::endl;
    
\end{verbatim}
\begin{verbatim}    
    for(n = 1; n < Nevents + 1; n++){

// read the n-th event from file
       ev.TakeEvent(Nfile,n,format,filename); 

// calculation of the variable t of the event
       t = 0.0;  
       x = (ev._particles()[0]._E()-ev._particles()[2]._E());
       t = -x*x;
       x = (ev._particles()[0]._px()-ev._particles()[2]._px());
       t = t+x*x;
       x = (ev._particles()[0]._py()-ev._particles()[2]._py());
       t = t+x*x;
       x = (ev._particles()[0]._pz()-ev._particles()[2]._pz());
       t = t+x*x; 
\end{verbatim}
\begin{verbatim}

// defines the t-bin
       i = 0;
       j = 0;
       while(tgrid[i] <= t && i < Nbins){
         j=i;
         i++;
       }
 
       result[j]++; // adds 1 to the bin
        
    }
        
// output        
       std::ofstream f;        
       f.open(ofilename, std::ofstream::out | std::ofstream::app);
        
       if (!f.is_open()) {
          std::cout << "Can not add to an *.res file!\n";   
        } 
        
       for(i = 0; i < Nbins; i++){ 
          result[i]=result[i]/Nevents;
          f << std::setw(9) << "{ " << std::setprecision(19) 
          << std::scientific << tbin[i] << ", " 
          << std::scientific << result[i] << " }, \n"; 
       }  
                
       f.close();
        
// time for calculations stop	//////////////////////////
       printf("clock stop = %d\n", std::clock());
       durationm = ( std::clock() - start ) / (double) CLOCKS_PER_SEC;
       std::cout<<"duration of the analizer = "<< durationm << '\n';
/////////////////////////////////////////////////////////
\end{verbatim}

Let us use any output \ttt{*.exdiff} file to extract events in the special
kinematical region.

In this example, we obtain the normalized $|t|$ probability distribution and put the result into a simple file. 

\subsection{For developers}
\label{ss:developers}

In this subsection we show how you can make modifications to perform another type of output format. For
this purpose we can use direct access to values of momenta and ids of all the particles. In the file
\ttt{ExDiff1.0/src/interface.cpp} you can find some comments:

\begin{verbatim}

// FOR DEVELOPERS ==================================================
// You can use directly all the parameters of particles in the event
// to make your own output or connection to your software
// instead of the string 

ev.AddToFile(outformat,outputfile);		

// use your own method with the following parameters of all particles ...
// i: 0 => ev._particles().size()-1 [i changes 0=>3 (2 to 2), ...
// ev._particles().size() = number of particles in the event
// ev._particles()[i]._ID() = ID of the	particle
// ev._particles()[i]._m() = mass of the	particle
// ev._particles()[i]._E() = energy of the	particle
// ev._particles()[i]._px() = px of the	particle
// ev._particles()[i]._py() = py of the	particle
// ev._particles()[i]._pz() = pz of the	particle	
// FOR DEVELOPERS ===================================================

\end{verbatim}

You could create your own class for output. In further versions of the generator we will add
\ttt{HEPMC} or other usual output.

%\subsection{Examples}
%\label{ss:examples}

\section*{Aknowledgements}

 Authors thanks to Simone Giani, Robert Ciesielski, Jan Kaspar, Gustavo Gil Da 
 Silveira, Michele Arneodo, Joao Varela and other members of the LHC diffractive community
 for useful discussions and help.
 
%----------references-----------------------------------------

\newpage
\section*{Index of basic Classes, Methods and Variables}
\addcontentsline{toc}{section}{Index of basic classes, methods and variables}

\boxsep

\noindent
\begin{minipage}[t]{\halfpagewid}
\begin{tabular*}{\halfpagewid}[t]{@{}l@{\extracolsep{\fill}}r@{}}
\ttt{ExDiff::NDarray } & \pageref{c:NDarray} \\
\ttt{ExDiff::NDarray::printC } & \pageref{m:NDarray.printC} \\
\ttt{ExDiff::NDarray::get } & \pageref{m:NDarray.get} \\
\ttt{ExDiff::NDarray::GetDimensions } & \pageref{m:NDarray.GetDimensions} \\
\ttt{ExDiff::NDarray::GetValues } & \pageref{m:NDarray.GetValues} \\
\ttt{ExDiff::NDarray::computeTotalSize } & \pageref{m:NDarray.computeTotalSize} \\
\ttt{ExDiff::NDarray::computeIndex } & \pageref{m:NDarray.computeIndex} \\
\ttt{ExDiff::NDarray::computeIndexes } & \pageref{m:NDarray.computeIndexes} \\
\ttt{ExDiff::NDarray::LoadFromFile } & \pageref{m:NDarray.LoadFromFile} \\
\ttt{ExDiff::NDarray::  dimensions } & \pageref{v:NDarray.dimensions} \\
\ttt{ExDiff::NDarray::  values } & \pageref{v:NDarray.values} \\
\ttt{ExDiff::NDgrid } & \pageref{c:NDgrid} \\
\ttt{ExDiff::NDgrid::GetVar } & \pageref{m:NDgrid.GetVar} \\
\ttt{ExDiff::NDgrid::GetDimensions } & \pageref{m:NDgrid.GetDimensions} \\
\ttt{ExDiff::NDgrid::GetValues } & \pageref{m:NDgrid.GetValues} \\
\ttt{ExDiff::NDgrid::get } & \pageref{m:NDgrid.get} \\
\ttt{ExDiff::NDgrid::computeIndex } & \pageref{m:NDgrid.computeIndex} \\
\ttt{ExDiff::NDgrid::computeTotalSize } & \pageref{m:NDgrid.computeTotalSize} \\
\ttt{ExDiff::NDgrid::printC } & \pageref{m:NDgrid.printC} \\
\ttt{ExDiff::NDgrid::LoadFromFile } & \pageref{m:NDgrid.LoadFromFile} \\
\ttt{ExDiff::NDgrid::  dimensions } & \pageref{v:NDgrid.dimensions} \\
\ttt{ExDiff::NDgrid::  values } & \pageref{v:NDgrid.values} \\
\ttt{ExDiff::iFunc } & \pageref{c:iFunc} \\
\ttt{ExDiff::iFunc::Calc } & \pageref{m:iFunc.Calc} \\
\ttt{ExDiff::iFunc::CheckPoint } & \pageref{m:iFunc.CheckPoint} \\
\ttt{ExDiff::iFunc::GetGrid } & \pageref{m:iFunc.GetGrid} \\
\ttt{ExDiff::iFunc::GetTab } & \pageref{m:iFunc.GetTab} \\
\ttt{ExDiff::iFunc::GetDimensions } & \pageref{m:iFunc.GetDimensions} \\
\ttt{ExDiff::iFunc::GetPoint } & \pageref{m:iFunc.GetPoint} \\
\ttt{ExDiff::iFunc::GetMinFunPoint } & \pageref{m:iFunc.GetMinFunPoint} \\
\ttt{ExDiff::iFunc::GetMaxFunPoint } & \pageref{m:iFunc.GetMaxFunPoint} \\
\ttt{ExDiff::iFunc::GetMinFun } & \pageref{m:iFunc.GetMinFun} \\
\ttt{ExDiff::iFunc::GetMaxFun } & \pageref{m:iFunc.GetMaxFun} \\
\ttt{ExDiff::iFunc::PrintVector } & \pageref{m:iFunc.PrintVector} \\
\ttt{ExDiff::iFunc::printC } & \pageref{m:iFunc.printC} \\
\ttt{ExDiff::iFunc::  dimensions } & \pageref{v:iFunc.dimensions} \\
\ttt{ExDiff::iFunc::  minvars } & \pageref{v:iFunc.minvars} \\
\ttt{ExDiff::iFunc::  maxvars } & \pageref{v:iFunc.maxvars} \\
\ttt{ExDiff::iFunc::  minfunpoint } & \pageref{v:iFunc.minfunpoint} \\
\ttt{ExDiff::iFunc::  maxfunpoint } & \pageref{v:iFunc.maxfunpoint} \\
\ttt{ExDiff::iFunc::  minfunvalue } & \pageref{v:iFunc.minfunvalue} \\
\ttt{ExDiff::iFunc::  maxfunvalue } & \pageref{v:iFunc.maxfunvalue} \\
\ttt{ExDiff::Generator} & \pageref{c:Generator} \\
\ttt{ExDiff::Generator::RNUM } & \pageref{m:Generator.RNUM} \\
\ttt{ExDiff::Generator::printC } & \pageref{m:Generator.printC} \\
\ttt{ExDiff::Generator::CalcMinus } & \pageref{m:Generator.CalcMinus} \\
\ttt{ExDiff::Generator::CalcFUNG } & \pageref{m:Generator.CalcFUNG} \\
\ttt{ExDiff::Generator::GetFUNGTOT } & \pageref{m:Generator.GetFUNGTOT} \\
\ttt{ExDiff::Generator::GenVar } & \pageref{m:Generator.GenVar} \\
%\ttt{ExDiff::Generator::EGenVarFast1} & \pageref{m:Generator.EGenVarFast1} \\
\ttt{ExDiff::Generator::IntVec } & \pageref{m:Generator.IntVec} \\
\ttt{ExDiff::Generator::Generate } & \pageref{m:Generator.Generate} \\
\ttt{ExDiff::Generator::  FUN, FUNG, FUNGTOT } & \pageref{v:Generator.FUN} \\
%\ttt{ExDiff::Generator::FUNG} & \pageref{v:Generator.FUNG} \\
%\ttt{ExDiff::Generator::FUNGTOT} & \pageref{v:Generator.FUNGTOT} \\
\end{tabular*}
\end{minipage}%
\newpage

\boxsep

\noindent
\begin{minipage}[t]{\halfpagewid}
	\begin{tabular*}{\halfpagewid}[t]{@{}l@{\extracolsep{\fill}}r@{}}
\ttt{ExDiff::Constants } & \pageref{c:Constants} \\
\ttt{ExDiff::Constants::Print } & \pageref{m:Constants.Print} \\
\ttt{ExDiff::Constants::\_mass } & \pageref{m:Constants.mass} \\
\ttt{ExDiff::Constants::\_dspin } & \pageref{m:Constants.dspin} \\
\ttt{ExDiff::Constants::  m\_p, m\_n, m\_pi0, m\_pi, m\_H, } & \pageref{v:Constants.mass} \\
\ttt{ExDiff::Constants::  m\_Gra, m\_R, m\_Glu, m\_Z, m\_W } & \pageref{v:Constants.mass} \\
\ttt{ExDiff::Constants::  alpha\_EM, Lam\_QCD} & \pageref{v:Constants.couplings} \\
\ttt{ExDiff::Constants::  VEV, M\_Pl} & \pageref{v:Constants.ewpars} \\
\ttt{ExDiff::Constants:: IDproton, ID1710, ID1950, } & \pageref{v:Constants.IDs} \\
\ttt{ExDiff::Constants:: IDpi0, IDpiplus, IDpiminus, IDdef } & \pageref{v:Constants.IDs} \\
\ttt{ExDiff::Particle } & \pageref{c:Particle} \\
\ttt{ExDiff::Particle::Print } & \pageref{m:Particle.Print} \\
\ttt{ExDiff::Particle::Rapidity } & \pageref{m:Particle.Rapidity} \\
\ttt{ExDiff::Particle::Pseudorapidity } & \pageref{m:Particle.Pseudorapidity} \\
\ttt{ExDiff::Particle::Theta } & \pageref{m:Particle.Theta} \\
\ttt{ExDiff::Particle::Phi } & \pageref{m:Particle.Phi} \\
\ttt{ExDiff::Particle::Lmult } & \pageref{m:Particle.Lmult} \\
\ttt{ExDiff::Particle::Lminus } & \pageref{m:Particle.Lminus} \\
\ttt{ExDiff::Particle::Lplus } & \pageref{m:Particle.Lplus} \\
\ttt{ExDiff::Particle::Lprint } & \pageref{m:Particle.Lprint} \\
\ttt{ExDiff::Particle::OnShell } & \pageref{m:Particle.OnShell} \\
\ttt{ExDiff::Particle::Reset } & \pageref{m:Particle.Reset} \\
\ttt{ExDiff::Particle::ResetP } & \pageref{m:Particle.ResetP} \\
\ttt{ExDiff::Particle::\_p} & \pageref{m:Particle.p} \\
\ttt{ExDiff::Particle::\_px,\_py,\_pz,\_E,\_m,\_pp,\_pT,\_p3D, } & \pageref{m:Particle.vars} \\
\ttt{ExDiff::Particle::\_DSpin,\_ID,\_Colour,\_Anticolour } & \pageref{m:Particle.vars} \\
\ttt{ExDiff::Particle:: p } & \pageref{v:Particle.p} \\
\ttt{ExDiff::Particle:: E, px, py, pz, m, pT, p3D } & \pageref{v:Particle.vars} \\
\ttt{ExDiff::Particle:: ID, DSpin, Colour, Anticolour } & \pageref{v:Particle.intvars} \\
\ttt{ExDiff::Event} & \pageref{c:Event} \\
\ttt{ExDiff::Event::Print } & \pageref{m:Event.Print} \\
\ttt{ExDiff::Event::\_particles } & \pageref{m:Event.particles} \\
\ttt{ExDiff::Event::AddToFile } & \pageref{m:Event.AddToFile} \\
\ttt{ExDiff::Event::TakeFromFile } & \pageref{m:Event.TakeFromFile} \\
\ttt{ExDiff::Event::TakeEvent } & \pageref{m:Event.TakeEvent} \\
\ttt{ExDiff::Event::CheckSum } & \pageref{m:Event.CheckSum} \\
\ttt{ExDiff::Event::  particles } & \pageref{v:Event.particles} \\
\ttt{ExDiff::KinCM } & \pageref{c:KinCM} \\
\ttt{ExDiff::KinCM::Print } & \pageref{m:KinCM.Print} \\
\ttt{ExDiff::KinCM::EvParticles } & \pageref{m:KinCM.EvParticles} \\
\ttt{ExDiff::KinCM::ResetVars } & \pageref{m:KinCM.ResetVars} \\
\ttt{ExDiff::KinCM::Phys } & \pageref{m:KinCM.Phys} \\
\ttt{ExDiff::KinCM::Transform\_CMpp\_to\_CMdd } & \pageref{m:KinCM.TransformPD} \\
\ttt{ExDiff::KinCM::Transform\_CMdd\_to\_CMpp } & \pageref{m:KinCM.TransformDP} \\
\ttt{ExDiff::KinCM::Transform\_CMpp\_to\_CMdd\_ } & \pageref{m:KinCM.TransformPD1} \\
\ttt{ExDiff::KinCM::Transform\_CMdd\_to\_CMpp\_ } & \pageref{m:KinCM.TransformDP1} \\
\ttt{ExDiff::KinCM::VtoP2 } & \pageref{m:KinCM.VtoP2} \\
\ttt{ExDiff::KinCM::VtoP3 } & \pageref{m:KinCM.VtoP3} \\
\ttt{ExDiff::KinCM::VtoP4 } & \pageref{m:KinCM.VtoP4} \\
\ttt{ExDiff::KinCM::PtoV2 } & \pageref{m:KinCM.PtoV2} \\
\ttt{ExDiff::KinCM::PtoV3 } & \pageref{m:KinCM.PtoV3} \\
\ttt{ExDiff::KinCM::PtoV4 } & \pageref{m:KinCM.PtoV4} \\
\ttt{ExDiff::KinCM::\_Y } & \pageref{m:KinCM.GetY} \\
\end{tabular*}
\end{minipage}%
\newpage

\boxsep

\noindent
\begin{minipage}[t]{\halfpagewid}
	\begin{tabular*}{\halfpagewid}[t]{@{}l@{\extracolsep{\fill}}r@{}}

\ttt{ExDiff::KinCM::  hadron1, hadron2 } & \pageref{v:KinCM.hadrons} \\
\ttt{ExDiff::KinCM::  particles } & \pageref{v:KinCM.particles} \\
\ttt{ExDiff::KinCM::  variables } & \pageref{v:KinCM.variables} \\
\ttt{ExDiff::KinCM::  sqs} & \pageref{v:KinCM.sqs} \\
\ttt{ExDiff::KinCM::  t, phi, DeltaT, } & \pageref{v:KinCM.vars22} \\
\ttt{ExDiff::KinCM::  Delta3D, theta, y, Delta } & \pageref{v:KinCM.vars22} \\
\ttt{ExDiff::KinCM::  t\_1, t\_2, phi\_12, phi\_1, } & \pageref{v:KinCM.vars23} \\
\ttt{ExDiff::KinCM::  xi\_1, xi\_2, y\_c, M\_T, M\_c } & \pageref{v:KinCM.vars23} \\
\ttt{ExDiff::KinCM::  DeltaT\_1, DeltaT\_2, } & \pageref{v:KinCM.vars23} \\
\ttt{ExDiff::KinCM::  Delta3D\_1, Delta3D\_2 } & \pageref{v:KinCM.vars23} \\
\ttt{ExDiff::KinCM::  Delta\_1, Delta\_2 } & \pageref{v:KinCM.vars23} \\
\ttt{ExDiff::KinCM::  t\_hat, s\_hat, Eta\_j, Theta\_j, Phi\_j } & \pageref{v:KinCM.vars24} \\
\ttt{ExDiff::KinCM::  flag} & \pageref{v:KinCM.flag} \\
\ttt{ExDiff::KinCM::  ID, masses, IDs } & \pageref{v:KinCM.varsinput} \\
\ttt{ExDiff::Interface } & \pageref{c:Interface} \\
\ttt{ExDiff::Interface::GeneratorInfo } & \pageref{m:Interface.GeneratorInfo} \\
\ttt{ExDiff::Interface::PrintPars } & \pageref{m:Interface.PrintPars} \\
\ttt{ExDiff::Interface::GetCard } & \pageref{m:Interface.GetCard} \\
\ttt{ExDiff::Interface::GenerateFile } & \pageref{m:Interface.GenerateFile} \\
\ttt{ExDiff::Interface::ResultFileName } & \pageref{m:Interface.ResultFileName} \\
\ttt{ExDiff::Interface::DatFileName } & \pageref{m:Interface.DatFileName} \\
\ttt{ExDiff::Interface::GridFileName } & \pageref{m:Interface.GridFileName} \\
\ttt{ExDiff::Interface::\_X} & \pageref{m:Interface.X} \\
\ttt{ExDiff::Interface::  IDauthors, IDprocess, IDenergy, } & \pageref{v:Interface.vars} \\
\ttt{ExDiff::Interface::  version, Nevents, } & \pageref{v:Interface.vars} \\
\ttt{ExDiff::Interface::  kinformat, outformat } & \pageref{v:Interface.vars} \\

\end{tabular*}
\end{minipage}%  

\end{document}